\documentclass[structabstract]{aa}
\usepackage{natbib}
\usepackage{graphicx}
\usepackage{txfonts}

\def\degree{^\circ}
\def\lya{Ly$-\alpha$}
\def\lyb{Ly$-\beta$}
\def\ion[#1 #2]{#1\,{\sc #2}}
\def\lamb[#1]{#1\,{\AA}}
\def\lambr[#1-#2]{{{#1}--{#2}\,{\AA}}}
\def\rat[#1 #2]{#1/#2}
\def\serts89{SERTS-89}

\def\tabul{\hbox{\raise 0.75pt\hbox{$\triangleleft$}}}
\def\ergs[#1]{#1 {ergs}~{cm$^{-2}$}\,{s$^{-1}$}\,{sr$^{-1}$}}
\def\dens[#1]{10$^{#1}$\hskip 1.5pt{cm$^{-3}$}}
\def\densr[#1 #2]{10$^{#1}$\hskip 1pt{--}\hskip .5pt{10$^{#2}$}\hskip 1.5pt{cm$^{-3}$}}
\def\fl[#1 #2]{{#1}$\pm${#2}}
\def\orb[#1 #2]{{$#1^{#2}$}}
\def\ls[#1 #2]{{$^{#1}${#2}}}
\def\tm[#1 #2 #3]{{$^{#1}${#2}$_{#3}$}}

\begin{document}
   \title{Solar transition region above sunspots}
   \author{H. Tian\inst{1, 2}
          \and W. Curdt\inst{1}
          \and L. Teriaca\inst{1}
          \and E. Landi\inst{3}
          \and E. Marsch\inst{1}}
   \institute{Max-Planck-Institut f\"ur Sonnensystemforschung,
   Max-Planck-Str. 2, 37191 Katlenburg-Lindau, Germany\\
   \email{tianhui924@gmail.com}
   \and School of Earth and Space Sciences, Peking University, China
   \and Naval Research Laboratory, Washington D.C., USA}

\abstract
   {}
   {We study the transition region (TR) properties above
   sunspots and the surrounding plage regions, by analyzing several sunspot
   reference spectra obtained by the SUMER (Solar Ultraviolet Measurements
   of Emitted Radiation) instrument in March 1999 and November 2006.}
   {We compare the SUMER spectra observed in the umbra, penumbra, plage, and sunspot
   plume regions. The hydrogen Lyman line profiles averaged in each of the four regions
   are presented. For the sunspot observed in 2006, the electron densities, differential
   emission measure (DEM), and filling factors of the TR plasma in the four
   regions are also investigated.}
   {The self-reversals of the hydrogen Lyman line profiles are almost absent in sunspots
   at different locations (at heliocentric angles of up to $49^\circ$) on the solar disk. In the
   sunspot plume, the Lyman lines are also not reversed, whilst the lower Lyman line profiles
   observed in the plage region are obviously reversed, a phenomenon found also in the normal
   quiet Sun. The TR densities of the umbra and plume are similar and one order of magnitude
   lower than those of the plage and penumbra. The DEM curve of the sunspot plume
   exhibits a peak centered at $\log(T/\rm{K})\sim5.45$, which exceeds the DEM
   of other regions by one to two orders of magnitude at these temperatures. We also find that
   more than 100 lines, which are very weak or not observed anywhere else on the Sun, are
   well observed by SUMER in the sunspot, especially in the sunspot plume. }
   {We suggest that the TR above sunspots is higher and probably more extended, and that the opacity of the
   hydrogen lines is much lower above sunspots, compared to the TR above plage regions. Our result indicates
   that the enhanced TR emission of the sunspot plume is probably caused by a large filling factor.
   The strongly enhanced emission at TR temperatures and the reduced continuum ensure that many
   normally weak TR lines are clearly distinctive in the spectra of sunspot plumes.}

   \keywords{Sun: UV radiation --
             Sun: transition region --
             Sun: sunspots --
             Line: profiles}
    \titlerunning{Solar transition region above sunspots}
    \authorrunning{H. Tian et al.}
   \maketitle
%
\section{Introduction}
The solar transition region (TR, between $\sim10^4~$K and $10^6~$K)
is the interface between the chromosphere and corona, where the
temperature and density change dramatically. Most of the TR emission
occurs in the VUV (vacuum ultraviolet) range of the electromagnetic
radiation \citep{Wilhelm07}. Thus, ultraviolet emission lines can
provide ample information about the magnetic structures and plasma
properties of the TR.

Earlier ultraviolet observations such as those made by the Naval
Research Laboratory (NRL) S082-B EUV spectrograph onboard the Skylab
space station \citep{BartoeEtal1977}, and NRL High-Resolution
Telescope Spectrograph (HRTS) flown on some rockets and Spacelab2
\citep{BruecknerEtal1977,BruecknerBartoe1983,BruecknerEtal1986},
provided much valuable information about the TR. These earlier
results were reviewed by \cite{Mariska92}.

Our knowledge of the TR has been enhanced greatly since the SUMER
(Solar Ultraviolet Measurements of Emitted Radiation) instrument
\citep{Wilhelm95,Lemaire97} onboard SOHO (Solar and Heliospheric
Observatory) began to observe in 1996. Because of its high spectral,
spatial, and temporal resolutions, and the wide wavelength coverage,
many more TR line profiles than in the past were obtained,
identified, and used in intensive studies \citep{Curdt01,Curdt04}.
In addition, the CDS (Coronal Diagnostic Spectrometer) instrument
\citep{Harrison1995} onboard SOHO has also increased significantly
our understanding of the TR structures and dynamics. For a review of
these recent progresses, we refer to \citet{Wilhelm07}.

\cite{Gabriel76} proposed a magnetic network model, in which the TR
emission originates in funnels diverging with height from the
underlying supergranular boundary. A decade later, \cite{Dowdy86}
proposed a modified model in which only a fraction of the network
flux opens, in the shape of funnels, into the corona, while the
remainder of the network is occupied by a population of low-lying
loops with lengths less than 10~Mm. Based on SUMER observations,
\cite{Peter01} suggested a new picture for the structure of the TR,
in which the funnels are either connected to the solar wind or just
the legs of large loops.

All the models mentioned above refer to the average structures in
the TR. However, the solar atmosphere is very inhomogeneous and
characterized by different large and small-scale structures.
Moreover, studies have shown that the TR is not thermally stratified
but strongly nonuniform and magnetically structured
\citep{Feldman83,Feldman87,Marsch06}. The SUMER instrument is well
suited to study the difference in TR structures in different regions
of the Sun. By combing the technique of magnetic field extrapolation
with SUMER observations, it has been found that the TR in coronal
holes is higher and more extended than in the quiet Sun
\citep{Tu05a,Tu05b,Tian08a,Tian08b}.

An active region is an area with an especially strong magnetic
field, where sunspots and plages are frequently formed. Our current
empirical knowledge and physical understanding of sunspots were
reviewed by \citet{Solanki03}. The sunspot spectra obtained by SUMER
reveal some distinct properties \citep{Curdt01}. For instance, some
spectral lines which are not observed in other areas of the Sun
stick out in the sunspot spectra. Moreover, in contrast to the quiet
Sun, the hydrogen Lyman line profiles in the sunspot are not
reversed. Until now, little has been done to understand these
phenomena.

Spectral lines with formation temperatures between $\sim10^5~$K and
$10^6~$K (upper-TR) often have significantly enhanced intensities at
locations overlying sunspot umbrae \citep{Foukal74}. These features
are usually termed sunspot plumes \citep{Foukal76}. A sunspot plume
usually has one end point anchored in the umbra and the other can
reach far from the sunspot. It is regarded as nothing more than the
common footpoints of several active region loops \citep{Dammasch08}.
Although sunspot plumes have been studied extensively, the reason
why the plume emission is so prominent at upper-TR temperatures is
still unknown.

Some work has been done to study the plasma properties of sunspot
plumes. The electron density, $\log(N_{\rm{e}}/\rm{cm}^{-3})$, of
the TR plasma in sunspot plumes is about 10 \citep{Doyle85,Doyle03}.
The sunspot plumes seem to be associated with downflows of TR plasma
\citep{Foukal76,Brynildsen2001,Marsch04,BrosiusLandi05,Dammasch08}.
The emission measure (EM) curve based on S-055 spectra of sunspot
plumes revealed two peaks at $\log\emph{(T/\rm{K})=}5.6$ and $6.1$,
respectively \citep{Noyes85}, while a more recent study showed that
the plume's differential emission measure (DEM) exhibited only one
peak centered at $\log\emph{(T/\rm{K})=}$5.6 or 5.8
\citep{BrosiusLandi05}. More work is needed to compare the
properties of sunspot plumes and other regions.

In this paper, we present a more complete analysis, by analyzing
more sunspot spectra obtained with SUMER at different locations of
the Sun , and compare the Lyman line (of main quantum number
\textit{n} higher than 2) profiles, electron densities, DEM curves,
and filling factors of the sunspot plume, umbra, penumbra, and the
surrounding plage regions. Our results have important physical
conclusions for the TR properties above the sunspot and the
surrounding plage region.
\section{Observation and data reduction}

We selected five reference spectra of two sunspots observed by
SUMER. One sunspot was observed at different locations of the solar
disk during March 16-19, 1999. The reference spectrum of the other
sunspot was obtained between 23:58 on November 13 and 02:52 of the
next day in 2006. The observational details are listed in
Table~\ref{table1}. The pointing in x and y is given in arcseconds
and refers to the slit center at the central time of the
observation. $\theta$ is the heliocentric angle.

\begin {table*}[]
\caption[]{Observational parameters of the 5 reference spectra of
sunspots. } \label{table1} \centering
\begin {tabular}{rccccccccc}
\hline\hline
year \vline & date & start time & end time & detector & slit & exposure time (s) & x ($^{\prime\prime}$) & y ($^{\prime\prime}$) & $\theta^{\rm{\circ}}$ \\
\hline
1999 \vline & March 16 & 13:09 & 16:14 & A & 6 & 90 & 44 & -351 & 22\\
 \vline & March 17 & 19:48 & 22:44 & B & 7 & 90 & 302 & -340 & 28\\
 \vline & March 18 & 17:35 & 20:30 & B & 7 & 90 & 479 & -360 & 38\\
 \vline & March 19 & 13:58 & 16:54 & B & 7 & 90 & 627 & -350 & 49\\
\hline
2006 \vline & November 13/14 & 23:58 & 02:52 & B & 7 & 90 &
-78 & -122 & 9 \\
\hline
\end {tabular}
\end {table*}

The standard procedures for correcting the SUMER data were applied,
including local-gain correction, dead-time correction, flat-field
correction, and image destretching.

In the earlier work of \citet{Curdt00}, the 1999 sunspot was
analyzed. However, only the reference spectrum obtained on March 18
was used. This spectrum was also used to produce the sunspot atlas,
and both the context image and slit position can be found in
\citet{Curdt01}. Here we extend the earlier work by analyzing four
reference spectra of this sunspot obtained at different solar
locations. Because of the solar rotation, this sunspot rotated from
the central meridian towards the limb, and its heliocentric angle
increases from $22 \degree$ to $49 \degree$. Thus, we can study the
center to limb variation in the Lyman line profiles of the sunspot,
as shown in Fig.~\ref{fig.1}.

The sunspot observed in 2006 was large in size, and several
reference spectra for it were obtained, all near the disk center.
For this reason, we analyze only one reference spectrum of this
sunspot. During this observation, TRACE (Transition Region and
Corona Explorer) obtained an image of the 1600~{\AA} passband at
00:32 on November 14, 2006. The sunspot image observed by TRACE is
shown in panel (A) of Fig.~\ref{fig.2}. Also shown there is the
approximate location of the SUMER slit. The SUMER instrument also
scanned this sunspot region by using several typical TR lines
(N~{\sc{iv}}, O~{\sc{iv}}, Si~{\sc{iv}}, O~{\sc{iii}}), and the
relevant results were published in \citet{Teriaca08}. Although the
scans were completed at different times, and a temporal variation
may have occurred in the sunspot region, their Fig.1 still provides
an impression of the TR emission from the sunspot and the
surrounding plage regions.

\section{Hydrogen Lyman line profiles}

Hydrogen is the most abundant element in the solar atmosphere and
its resonance lines play an important role in the overall radiative
energy transport of the Sun \citep{Fontenla88}.

Line profiles of the full hydrogen Lyman series can be acquired by
SUMER at high spectral resolution. It has been found that the
average profiles for \lyb~$(n=$~2) through {Ly$-\epsilon$~($n=$~5)}
are self-reversed and stronger in the red horns, while the higher H
Lyman series lines (from {Ly$-\zeta$~to {Ly$-\lambda$, ($n=6,\cdots,
11$)}) are flat-topped \citep{Warren98,Xia03,Xia04}. Higher Lyman
lines obtained near the limb were analysed by \citet{Marsch99} and
\citet{Marsch00}, in which the authors found an increase in the line
width with decreasing main quantum number and an unexpectedly flat
hydrogen-temperature gradient. The \lya~profiles in the quiet Sun
were obtained by SUMER through several non-routine observations, by
closing the aperture door of SUMER to reduce the incoming photon
flux to a 20\%-level \citep{Curdt08, Tian09}. The average
\lya~profile was found to be strongly reversed and have a stronger
blue horn. It is believed that the opposite asymmetries in the
average profiles of \lya~and higher Lyman lines are probably caused
by the combined effect of flows in the different layers of the solar
atmosphere and opacity differences of the lines \citep{Gunar08,
Tian09}.

The Lyman series are important for diagnosing the variation in the
thermodynamic conditions in prominences \citep{Vial07}. Prominence
thread models including multi-level non-LTE transfer calculations
have shown that the profiles of Lyman lines are more reversed when
seen across than along the magnetic field lines \citep{Heinzel05}.
This behaviour was confirmed in a prominence observation by
\citet{Schmieder07}.

\begin{figure*}
\resizebox{\hsize}{!}{\includegraphics{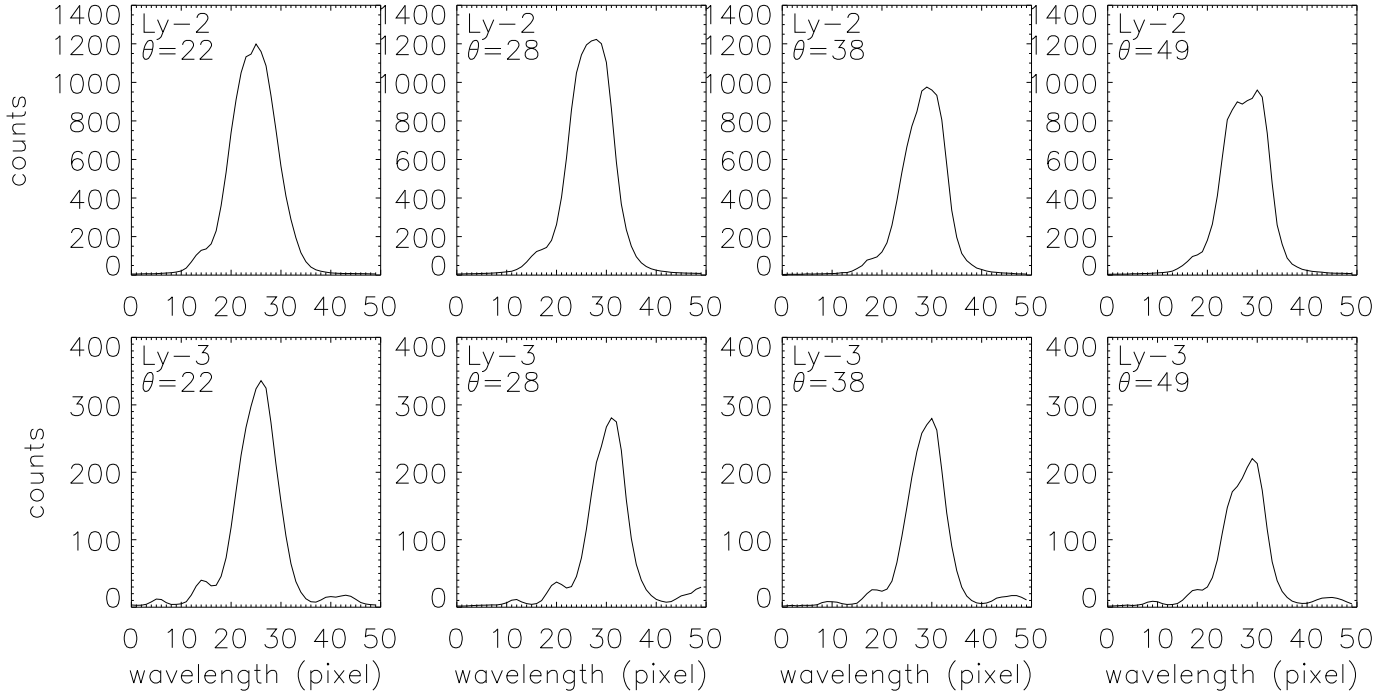}}
\caption{Averaged profiles of Ly-2 (upper panels) and Ly-3 (lower
panels) in the sunspot observed on 16th ($\theta=22^\circ$), 17th
($\theta=28^\circ$), 18th ($\theta=38^\circ$), and 19th
($\theta=49^\circ$) of March in 1999. $\theta$ is the heliocentric
angle.} \label{fig.1}
\end{figure*}

In sunspot regions, the Lyman line profiles exhibit properties that
are different from the average profiles. The sunspot atlas of SUMER
reveals that the Lyman line profiles observed in the sunspot are
almost not reversed \citep{Curdt01}. However, the authors did not
mention this phenomenon in that paper. Here we extend this earlier
work by analyzing four reference spectra of this sunspot obtained at
different solar locations. Since the magnetic field lines in
sunspots are almost vertical, we can use the 4 spectra to study the
properties of the Lyman line profiles in sunspots and the dependence
of their self-reversals on the orientation of field lines.

The Lyman line profiles in the four spectra are all not or only
slightly reversed, similar to those in off-limb coronal hole
observations \citep{Marsch00}. This result suggests that the TR
plasma of the sunspot is almost optically thin, regardless of the
location where the sunspot is observed. Since it is known that the
higher order Lyman lines are optically thinner than lower order
Lyman lines, radiative transfer effects, if present, should be more
pronounced in lower order Lyman lines than in higher order Lyman
lines. Thus, we present only the Ly-2 and Ly-3 profiles averaged
over the sunspot (umbra) portion of the slit in Fig.~\ref{fig.1}.
Note that there are some blends with the Lyman lines
\citep{Curdt01}. From an inspection of Fig.~\ref{fig.1}, it seems
that profiles begin to reverse when the vertical field lines of the
sunspot make an angle of about $49^\circ$ with respect to the line
of sight. This phenomenon seems to be similar to the observational
result for prominences, namely that the Lyman profiles are more
reversed when seen across the field lines than along the field lines
\citep{Schmieder07}. However, from $\theta=22^\circ$ to
$\theta=38^\circ$, it is difficult to say whether or not there is a
trend in the shapes of the top parts of the profiles.

\begin{figure*}
\sidecaption
\includegraphics[width=12.5cm]{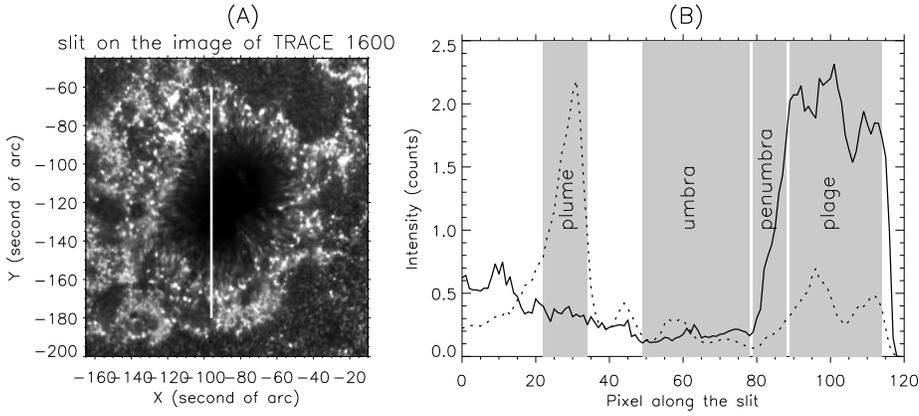}
\caption{The sunspot observed on 14th of November, 2006. (A) The
SUMER slit is located on the image of the 1600~{\AA} passband of
TRACE. The TRACE image was obtained at 00:32 on November 14, 2006.
(B) The intensities of continuum around 1045~{\AA} (solid line) and
O~{\sc{vi}} (1031.9~{\AA}, dotted line) along the slit. Four
segments corresponding to the plage, penumbra, umbra, and sunspot
plume are marked in grey. The O~{\sc{vi}} intensity has been divided
by 600.} \label{fig.2}
\end{figure*}

\begin{figure*}
\resizebox{\hsize}{!}{\includegraphics{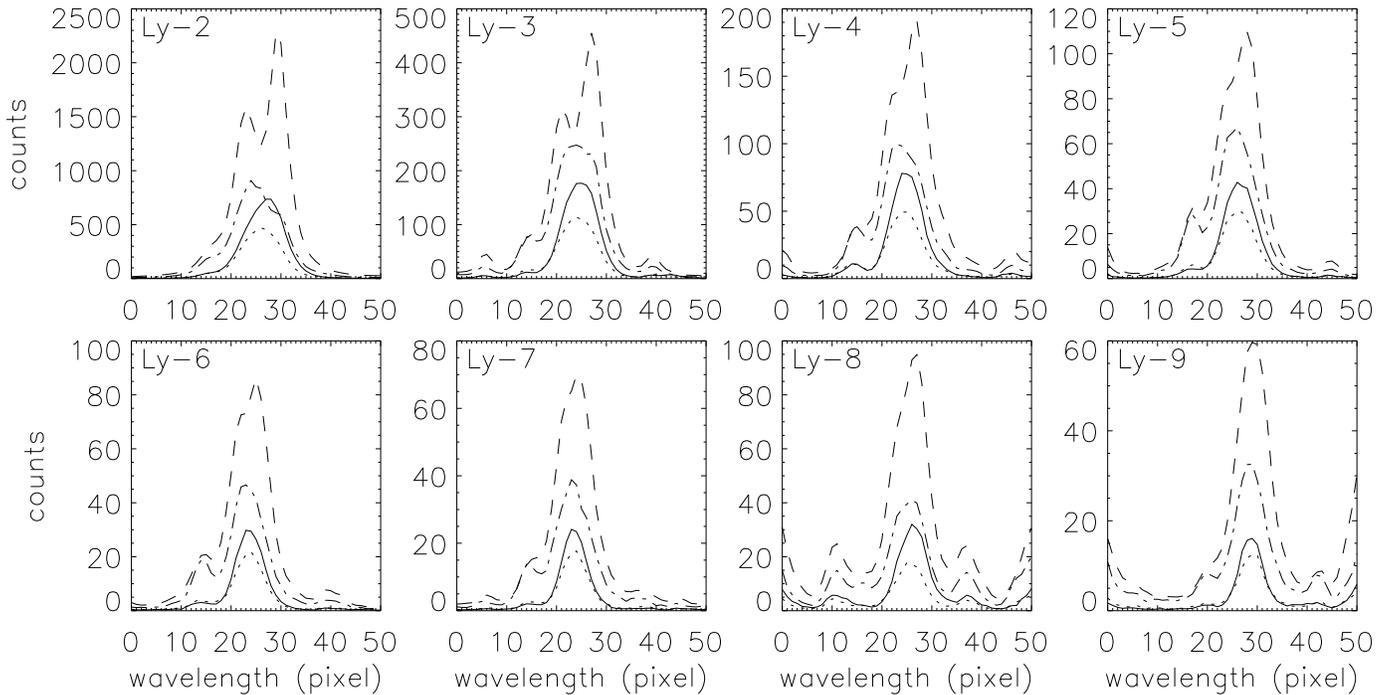}} \caption{The
dashed, dot-dashed, dotted, and solid lines represent the average
profiles of the Lyman lines (from Lyman-2 to Lyman-9) observed in
the plage, penumbra, umbra, and plume for the 2006 data set. Note
that the profiles of the umbra were multiplied by a factor of two. }
\label{fig.3}
\end{figure*}

The sunspot observed in 2006 is a large one, and the SUMER slit also
caught part of a sunspot plume and the surrounding plage region.
Figure~\ref{fig.2} shows the intensities of the continuum around
1045~{\AA} and of the O~{\sc{vi}} (1031.9~{\AA}) line along the
slit, and the resulting curves clearly show the umbra, penumbra, and
plage regions as well as part of the sunspot plume. In this paper,
the term intensity is regarded as being equivalent to radiance. The
sunspot plume shows enhanced radiance in typical TR lines, which
corresponds to the slit segment with the highest intensities of
O~{\sc{vi}} in our reference spectra. We selected four portions of
the slit, which are marked in grey in Fig.~\ref{fig.2} and
correspond to the umbra, penumbra, plage, and plume locations. The
average Lyman line profiles (from Ly-2 to Ly-9) of the four parts
are presented in Fig.~\ref{fig.3}. For the blends with the Lyman
lines, we refer to \citet{Curdt01}.

Our results show that the self-reversals of all the hydrogen Lyman
line profiles are absent in the umbra and plume regions. The lower
Lyman line profiles observed in the plage region are obviously
reversed, a phenomenon similar to the normal quiet Sun \citep[one
can refer to the atlas presented in][]{Curdt01}. The profiles in the
penumbra are not so peaked as those in the umbra and plume, and not
so reversed as those in the plage. We also note that the Lyman line
profiles observed in the plage have a very strong red-horn asymmetry
(the red horn is much stronger than the blue horn), which might be
at least partly caused by the strong red shift observed in the TR.
The asymmetries of the lower Lyman line profiles in the penumbra
seem to be opposite to those of the plage, which might indicate a
different pattern of flows in the upper atmosphere of the two
regions.

Unfortunately, so far SUMER has not been used to complete
\lya~observations of the sunspot. We note that \citet{Fontenla88}
presented a \lya~profile (see Fig.12 in the paper) of a sunspot
observed by the Ultraviolet Spectrometer and Polarimeter on the
\emph{SMM} (Solar Maximum Mission) spacecraft. As mentioned in that
paper, the emission suffered from geocoronal absorption. After
correcting for this effect, their \lya~profile seems to be
flat-topped, which is very different from the strongly reversed
\lya~profiles in the quiet Sun \citep{Curdt08, Tian09}.

The above results indicate that the opacity is strongly reduced
above the sunspot, with respect to the surrounding plage region. By
analysing the emission lines of H$_{2}$ in the sunspot as well as
the quiet Sun, \citet{Jordan78} and \citet{Bartoe79} concluded that
the opacity over the sunspot is about an order of magnitude lower
than in the quiet Sun. The different opacities above sunspots and
plages are confirmed by the intensity ratio between \lya~and \lyb~,
which is about 200 in the sunspot umbra and 130 in the plage. In the
sunspot, the profiles are only weakly absorbed. In the plage, the
opacity is higher and thus the absorption is enhanced. Since
\lya~has a larger opacity, its absorption will be stronger than
\lyb, which will lead to a lower observed value of the intensity
ratio. However, we note that the \lya~line was recorded on the
attenuator, so the measurement of its intensity is highly uncertain
and we do not consider it in detail here.

Since the Lyman line profiles in the plage are quite similar to
those observed in the normal quiet Sun, we may expect a similar
opacity in both regions. This similarity might be the result of the
similar magnetic structures. In both regions, magnetic loops
reaching different heights are the dominant structures. The emission
sources of Lyman lines are located mainly in loops that reach the
layers of chromosphere and TR. When the chromospheric Lyman line
photons travel through the upper atmosphere, they will be initially
absorbed by the hydrogen atoms in upper-chromospheric and TR loops,
and further absorbed in coronal loops. Since the density decreases
and the temperature increases with height above the temperature
minimum, there are more hydrogen atoms, and the absorption is much
stronger in the upper chromosphere and lower-TR layers than at
higher layers. In the plage region and quiet Sun, the Lyman line
emission originates by a large fraction in the chromosphere and are
strongly absorbed in the upper chromosphere and lower TR, leading to
a strong absorption at the center of the Lyman line profiles.

The scenario seems to differ in sunspot regions. The almost
Gaussian-shaped profiles suggest a weak absorption of the Lyman line
emission. This observational result seems to favor a scenario where
there is less chromospheric plasma above sunspots, which might be
the case because the sunspot is much cooler than the surrounding
regions. In this case, the ratio of the chromospheric to TR
contributions of the Lyman line radiation is lower in sunspots than
in plage regions. As a result, the chromospheric emission will only
be weakly absorbed because there is little absorbing material in the
upper chromosphere. Moreover, the absorption of the TR emission is
weak because of the lower density above the emission sources.

Sunspot plumes often show greatly enhanced emission at upper-TR
temperatures ($\sim10^5$ and $10^6~$K), while at lower or higher
temperatures the emission is rather weak
\citep[e.g.,][]{Foukal74,BrosiusLandi05}. Sunspot plumes are
frequently reported to be associated with TR downflows
\citep[e.g.,][]{Marsch04,Dammasch08}. Observational results seem to
reveal that plasmas at TR temperatures dominate in plumes. Thus, the
ratio of TR to chromospheric contributions to the Lyman line
emissions may be significantly higher in plumes. These plume loops
are large in size (and far reaching) and the overlying corona is
insufficiently dense to cause an obvious dip at the center of the
Lyman line profiles.

The above explanations are based mainly on a static stratified
atmosphere. However, it is well known that the real solar
atmosphere, especially the TR over sunspots, is rather inhomogeneous
and dynamic \citep{Brynildsen1999a,Brynildsen1999b}. It has been
found that the Lyman line profiles can be modified by TR flows in
the quiet Sun \citep{Curdt08, Tian09}. In sunspot regions, both
significant upflows and downflows have been frequently reported
\citep[e.g.,][]{Kjeldseth-Moe1988,Teriaca08}. Thus, we should not
exclude the possibility that additional effects, such as the
presence of velocities and radiation penetrating from the sides in
the inhomogeneous plasma, may alter the source functions of the
Lyman lines and also affect the Lyman line profiles.

\section{Electron densities}

Based on the assumption that the plasma is in ionization
equilibrium, one can calculate the electron density from line-ratio
observations \citep[for a review, see][]{Xia03,Wilhelm04}.
Density-sensitive line pairs usually include two emission lines
within the same ion. In the de-excitation process, the relative
importance of radiative decay to collisional de-excitation is
different for the two lines.

\begin{table*}[]
\caption[]{Electron density ($\log(N_{\rm{e}}/\rm{cm}^{-3}$))
measurements, for the sunspot observed in 2006. } \label{table2}
\centering
\begin{tabular}{lrcllll}
Ion & Wavelength pair (${\AA}$) & $T_{\rm{f}}~(\log$~K) & Umbra & Penumbra & Plage & Plume \\
\hline\hline
\ion[Si iii]  & 1301.16/1298.96 & 4.68 & 11.1$^{+0.5}_{-0.4}$ & $> 11.0$ & & 11.1$^{+0.5}_{-0.4}$ \\
\ion[C iii]   & 1175.98/1175.24 & 4.84 & $> 9.3$ & $> 9.3$ & $> 9.2$ & $> 9.4$ \\
\ion[O iv]    & 1399.77/1401.16 & 5.18 & 10.2$^{+0.4}_{-1.2}$ & 11.0$^{+0.7}_{-0.5}$ & 10.8$^{+0.5}_{-0.5}$ & 10.3$^{+0.4}_{-0.3}$ \\
\ion[O iv]    & 1407.39/1401.16 & 5.18 & 10.3$^{+0.4}_{-0.8}$ & 10.9$^{+0.5}_{-0.4}$ & 11.0$^{+0.6}_{-0.4}$ & 10.4$^{+0.4}_{-0.7}$ \\
\ion[O v]     & 758.68/761.13   & 5.37 & 10.1$^{+0.1}_{-0.2}$ & 10.9$^{+0.3}_{-0.2}$ & 11.0$^{+0.3}_{-0.3}$ & 10.1$^{+0.2}_{-0.1}$ \\
\ion[O v]     & 759.43/761.13   & 5.37 & 10.0$^{+0.2}_{-0.1}$ & 11.0$^{+0.5}_{-0.2}$ & 11.0$^{+0.5}_{-0.2}$ & 10.1$^{+0.1}_{-0.2}$ \\
\ion[O v]     & 760.43/761.13   & 5.37 & 10.0$^{+0.2}_{-0.2}$ & 11.2$^{+0.6}_{-0.3}$ & 11.0$^{+0.3}_{-0.3}$ & 10.1$^{+0.2}_{-0.2}$ \\
\ion[O v]     & 761.99/761.13   & 5.37 & 10.0$^{+0.2}_{-0.1}$ & 10.9$^{+0.4}_{-0.2}$ & 11.0$^{+0.4}_{-0.2}$ & 10.0$^{+0.2}_{-0.1}$ \\
\ion[Mg viii] & 769.38/782.34   & 5.90 & $< 12.0$ & & & $< 11.9$ \\
\ion[Fe xii]  & 1349.43/1241.95 & 6.13 & 9.3$^{+1.3}_{-1.3}$ & $< 10.1$ & $< 10.2$ & $< 7.6$ \\
\hline
\end{tabular}
\end{table*}

There are several density-sensitive line pairs in the SUMER spectral
range \citep{Wilhelm95}. However, some of these lines are either
severely blended with other lines and difficult to decompose, or too
weak for a reliable calculation. We selected as many lines as
possible to calculate the densities for the sunspot observed in
2006. All of these line pairs have a weak dependence on the electron
temperature. Fortunately, the few density-sensitive lines are close
in wavelength, so that the effects of temporal variations on the
measured densities are minimized. By applying the method of single-
or multi-Gaussian fitting to the profiles of these lines averaged in
the plage, umbra, penumbra, and plume, we were able to obtain the
intensities. The theoretical relations between intensity ratios of
line pairs and electron densities were taken from the CHIANTI data
base \citep{Dere97,Landi06}. The details of the line pairs and
density calculations are listed in Table~\ref{table2}. Here
$T_{\rm{f}}$ represents the formation temperature of the ion. We
attempted to measure the electron density in all temperature
regimes, but accurate measurements could only be carried out for the
lower-temperature region; the available coronal lines allow us to
derive only broad estimates. We assume a 15\% uncertainty in the
determination of the line intensities. This error finally propagates
into the uncertainties for the densities listed in
Table~\ref{table2}.

The most accurate measurements are obtained in the transition region
from \ion[O iv] and \ion[O v]. The O~{\sc{iv}}~1401.16~{\AA} line is
blended by the chromospheric line S~{\sc{i}}~1401.51~{\AA}, and the
O~{\sc{iv}}~1407.39~{\AA} line is blended with two second order
O~{\sc{iii}}~lines (703.845~{\AA}, 703.85~{\AA}). However, both of
these blends can be resolved by a two-component Gaussian fitting,
and thus we are able to obtain reliable intensities of the
O~{\sc{iv}}~1407.39~{\AA} and O~{\sc{iv}}~1401.16~{\AA} lines.

The electron densities derived by using O~{\sc{v}} line pairs show a
general consistency. In the plage and penumbra regions, the value of
$\log(N_{\rm{e}}/\rm{cm}^{-3})$ is around 11, which is one order of
magnitude larger than in the umbra and plume regions. The derived
density in the sunspot plume here is consistent with those derived
by \citet{Doyle85} and \citet{Doyle03}. In \citet{Doyle85}, the
authors used the ratio of O~{\sc{v}}~760~{\AA}/630~{\AA} obtained by
the S-055 EUV spectrometer onboard \emph{Skylab} and derived a
density of $\log(N_{\rm{e}}/\rm{cm}^{-3})=10$. \citet{Doyle03}
measured the density of the sunspot plumes observed by SUMER on
March 18, 1999. By using the same O~{\sc{v}} line pairs as we did,
they obtained a density of $\log(N_{\rm{e}}/\rm{cm}^{-3})=9.9$. The
O~{\sc{v}}~759.43 line is suggested to be blended by
S~{\sc{iv}}~759.34 \citep{Curdt01}. However, the measurement of
\citet{Doyle03} indicates that this blend should not be significant.
Our result confirms this finding.

The average umbra and plume densities derived by using the
O~{\sc{iv}} line pairs seem to be higher than in the O~{\sc{v}}
results. However, the differences are within the uncertainties. We
found that the intensity of the plume part seen in O~{\sc{iv}} is
not as strong as that seen in O~{\sc{v}}, which indicates that the
strongest part of the plume emission might not have been caught by
the slit, or that the plume was less prominent at the observation
time of the O~{\sc{iv}} line pairs. By using the ratio of
O~{\sc{iv}}~625~{\AA}/790~{\AA} from the S-055 sunspot plume
spectrum, \citet{Doyle85} derived a density of
$\log(N_{\rm{e}}/\rm{cm}^{-3})=10.3$, which is very close to our
plume density at the same temperature. However, by using the line
pair of O~{\sc{iv}}~625.8~{\AA}/554.5~{\AA} measured by CDS,
\citet{BrosiusLandi05} derived relatively low values
($\log(N_{\rm{e}}/\rm{cm}^{-3})=9.6\sim9.7$) of plume density.
However, the line of O~{\sc{iv}}~625.8~{\AA} was rather weak and
blended with Mg~{\sc{x}}~624.9~{\AA} in the red wing, so the derived
densities were highly uncertain.

The \ion[C iii] and \ion[Si iii] measurements only provide lower
limits to the electron density. The \ion[C iii] values agree with
the TR values from the oxygen ions, while \ion[Si iii] infers a
higher density than the oxygen ions in the umbra and plume. This
might suggest the presence of a significant density gradient in the
umbra and in the plume from the chromosphere to the lower TR, which
would indicate a decrease in the plasma pressure, if we calculate
the pressure to be $p=N_{\rm{e}}~T_{\rm{f}}$. However, possible
strong temperature gradients on one side, and opacity effects in the
\ion[C iii] and \ion[Si iii] on the other, cast huge uncertainties
on the presence of a density and pressure gradient in the umbra and
the plume.

In the corona, \ion[Mg viii] and \ion[Fe xii] provide only very
rough estimates of or upper limits to the electron densities, which
are consistent with almost any value measured in the corona in
quiescent conditions. In this case, our SUMER dataset is unable to
provide significant constraints.

The umbra and plume densities derived here are similar to, or
slightly larger than, the density of the normal quiet Sun, which has
an upper limit of $\log(N_{\rm{e}}/\rm{cm}^{-3})=9.87$ at
$\log(T/\rm{K})=5.25$ in \cite{GriffithsEtal1999}. A similar result
is also obtained by \citet{BrosiusLandi05}, in which the densities
of plumes and quiet Sun areas are estimated to be around
$\log(N_{\rm{e}}/\rm{cm}^{-3})=9.6\sim9.7$ and
$\log(N_{\rm{e}}/\rm{cm}^{-3})=9.4$, respectively, by using the same
line pair of O~{\sc{iv}}~625.8~{\AA}/554.5~{\AA} as observed by the
CDS instrument.

By analyzing the EUV data obtained by the Harvard spectrometer on
the Apollo telescope mount, \citet{Foukal74} measured a significant
decrease in the gas density of the umbra relative to the surrounding
plage. Here we confirm this result by finding that the densities of
the umbra and plume at TR temperatures are about a factor of 10
lower than of the plage. Our measurements seem to indicate that the
sunspot plasma emitting at TR temperatures is higher and probably
more extended than in the surrounding plage region. Since the
density of the solar atmosphere decreases almost exponentially with
height above the photosphere, this scenario naturally leads to a
much lower density in the TR above sunspots, as compared to the TR
above the surrounding plage. This scenario also implies that the
sunspot TR temperature is much lower than the surrounding
temperature at the same heights. Our scenario predicts a temperature
structure of sunspots as that proposed by \cite{Nicolas1982}. Our
conclusion is also consistent with that of \cite{Guo09}, in which
the authors suggested that stronger magnetic fields correspond to
higher formation heights of VUV lines.

We note that this TR scenario is similar to that for a coronal hole,
since the TR in the coronal hole is also found to be higher and more
extended than the TR in the quiet Sun
\citep{Tu05a,Tu05b,Tian08a,Tian08b}. The average \lyb~and Ly-3
profiles in the coronal hole atlas are not reversed, a phenomenon
similar to the sunspot, whilst they are obviously reversed in the
quiet Sun \citep{Curdt01}. Moreover, solar wind flows out along
magnetic funnels in coronal holes \citep{Tu05a,Esser05}, and the
signature of upflows associated with open field lines was also found
in sunspots \citep{Marsch04}. These results infer a similarity
between the properties of the TR above sunspots/plages and the TR in
CH/QS (coronal hole and quiet Sun).


It was suggested that the downflow of TR plasma is essential to the
existence of plumes \citep{Brynildsen2001,Foukal76}. However, it
remains debated whether this is achieved by siphon flows along far
reaching loops \citep{Brynildsen2001,Doyle03,Brosius05} or by
cooling and condensing coronal plasma falling downward along the
``cold surface" of the plume
\citep{Noyes85,BrosiusLandi05,Dammasch08}. A siphon flow is possibly
driven by a strong asymmetric heating and the resulting pressure
difference in two loop legs
\citep{McClymontCraig1987,Mariska1988,SpadaroEtal1991}. This type of
flow may cause a blue shift in one leg and red shift in the other.
\citet{Brynildsen2001} concluded that the inflow of plasma at TR
temperatures from locations well outside the sunspots is a necessary
requirement for the sunspot plume to occur. This hypothesis seems to
be supported by the observational result of \citet{Brosius05}, in
which a significantly higher density
($\log(N_{\rm{e}}/\rm{cm}^{-3})=9.8$, compared to
$\log(N_{\rm{e}}/\rm{cm}^{-3})=8.9$ in the plume, in an upflow
region outside the sunspot was observed by CDS. \citet{Doyle03} also
measured higher densities (with $\log(N_{\rm{e}}/\rm{cm}^{-3})$
ranging from 10.20 to 10.45) in nearby quiet regions (their $QS_{1}$
might be a plage region) than in the plume
($\log(N_{\rm{e}}/\rm{cm}^{-3})\sim9.9$), and suggested that the gas
pressure difference might be sufficient to drive siphon flows from
outside the spot into the umbra. Our measured densities in different
regions also seem to support this idea. It is generally believed
that sunspot plumes are associated with loops in which one leg is
anchored in the umbra and the other anchored outside the sunspot,
probably in the plage region. Our measurement shows that the
densities at TR temperatures are one order of magnitude lower in the
umbra and plume than in the plage. This pressure difference is
certainly sufficient to initiate a siphon flow in the loop.

\section{Differential emission measure}

\subsection{Method}

The plasma in the upper solar atmosphere can usually be assumed to
be optically thin, although this assumption does not hold for the
coldest ions observed in the SUMER spectra, such as neutrals and
singly ionized species. In the optically thin case, the radiance of
a line can be written as
\begin{eqnarray}
I_{ji} =
\frac{1}{4\pi}\int_{}^{}G_{ji}{\left({T,N_{\rm{e}}}\right)}N_{\rm{e}}^{2}dh,
\label{intensityintegral2}
\end{eqnarray}

\noindent where $N_{\rm{e}}$ is the electron density, $h$ is the
height of the emitting volume along the line of sight, and
$G_{ji}{\left({T,N_{\rm{e}}}\right)}$ is the line {\em contribution
function}.




The contribution function includes all the atomic physics involved
in the process of line formation, and can be computed as a function
of temperature and density using spectral codes that include all the
atomic parameters necessary to calculate the contribution function.
In the present work we will use version 5.2.1 of the CHIANTI
database \citep{Dere97,Landi06}, the ion fraction dataset of
\citet{Mazzotta98}, and the photospheric abundances of
\citet{Grevesse98}.

The $N_{\rm{e}}^2dh$ term beneath the sign of the integral on the
right-hand side of Eq~(\ref{intensityintegral2}) includes plasma
conditions involved in the process of line formation. It is an
important quantity when studying the thermal structure of the solar
corona and comparing predictions from theoretical models with
observations. When the plasma is multi-thermal and there is a
continuous relationship between the amount of plasma and
temperature, then Eq~(\ref{intensityintegral2}) can be rewritten by
defining the {\em Differential Emission Measure} (DEM) function
$\varphi{\left({T}\right)}$ as
\begin{eqnarray}
\varphi{\left({T}\right)} & = & N_{\rm{e}}^2\frac{dh}{dT}, \\
\label{dem2} I_{ji} & = &
\frac{1}{4\pi}\int_{}^{}G_{ji}{\left({T,N_{\rm{e}}}\right)}\varphi{\left({T}\right)}dT,
\label{dem1}
\end{eqnarray}

\noindent where the DEM indicates the amount of material in the
plasma as a function of temperature. Several methods have been
developed to determine $\varphi{\left({T}\right)}$ from a set of
observed lines. Reviews of the main examples of these methods can be
found in \citet{Harrison92} and \citet{Phillips08}. In the present
work, we use the iterative technique developed by \citet{Landi97}.
In this technique, an initial, arbitrary
$\varphi_0{\left({T}\right)}$ curve is first assumed. Corrections to
this curve are calculated by evaluating the ratio of observed line
fluxes to theoretical values predicted using the
$\varphi_0{\left({T}\right)}$ curve. Each correction
$\omega_0{\left({T_{\rm{eff}}}\right)}$ is associated with an
effective temperature $T_{\rm{eff}}$ defined as

\begin{equation}
\log T_{\rm{eff}} =
{{\int{G{\left({T,N_{\rm{e}}}\right)}\varphi_0{\left({T}\right)}\log
T~dT}}\over
{\int{G{\left({T,N_{\rm{e}}}\right)}\varphi_0{\left({T}\right)}
dT}}}. \label{teff}
\end{equation}

\noindent A new $\varphi_1{\left({T}\right)}$ curve is determined by
first calculating the values of the new
$\varphi_1{\left({T}\right)}$ curve at the temperatures
$T_{\rm{eff}}$ to be

\begin{equation}
\varphi_1{\left({T_{\rm{eff}}}\right)}=\varphi_0{\left({T_{\rm{eff}}}\right)}\times
\omega_0{\left({T_{\rm{eff}}}\right)},
\end{equation}

\noindent and then interpolating the results over temperatures to
provide a continuous function. The resulting
$\varphi_1{\left({T}\right)}$ curve is then used as an initial DEM
curve in the next iteration to calculate new corrections. The
procedure converges when the corrections to the $i-th$ DEM curve are
all unity within the uncertainties. This technique reaches the final
solution in just a few iterations, and the final result is
independent of the initial arbitrary $\varphi_0{\left({T}\right)}$
curve.

\subsection{DEM curves}

We adopted an electron density of
$\log(N_{\rm{e}}/\rm{cm}^{-3})=10.0$ for the umbra and the plume,
and $\log(N_{\rm{e}}/\rm{cm}^{-3})=11.0$ for the penumbra and plage,
to calculate the contribution functions to be used for DEM
diagnostics. The DEM curves determined using the spectral lines
listed in Table~5 are shown in Fig.~\ref{fig4}. To obtain those
curves, it was necessary to complete an additional selection of the
spectral lines to be used, as well as to modify the adopted element
abundances.

When we applied the DEM diagnostic technique to {\em all} the listed
lines at once, very large disagreements were found between lines of
the same ion, and between lines of different ions with similar $\log
T_{\rm{eff}}$. These disagreements occurred in all four regions, and
made it impossible to calculate a DEM curve for each region because
1) the correction curves $\omega{\left({T}\right)}$ never converged
to unity and showed an oscillatory behavior, and 2) even when they
were close to unity, the disagreements between different ions and
between lines of the same ion were so large that the resulting DEM
curve was meaningless.

The causes of these problems are probably numerous: atomic physics
problems in the level population calculation within each ion;
inaccurate ion abundances; opacity effects (for the coldest lines in
the dataset); non-photospheric element abundances; or temporal
variability. Opacity effects can be ruled out for all the listed
lines, except for the strongest lines belonging to the ions formed
at lowest temperatures. The intensities of all other cold lines are
too weak to provide significant self-absorption. Moreover,
\citet{Brooks00} ruled out opacity effects for many of the coldest
ions, although they showed that there may be significant opacity
effects for \ion[C ii] and \ion[C iii]. Atomic physics problems can
be ruled out as the discrepancies between lines of the same ions are
much larger than any effect of inaccurate collisional or radiative
excitation rates. The same can be said for ion abundances, unless
non-equilibrium conditions are present in the emitting plasma. The
most likely cause of the problems that we encountered is the
temporal variability of the plasma. Any temporal changes in the
physical properties of the plasma directly affect our DEM results
because the SUMER data were taken over several hours, and the
transition region plasma is known to be changing even within a few
minutes.

To minimize the effects of temporal variability, we restricted our
analysis to lines that were emitted within about 20 minutes. Since
SUMER scans its wavelength range by shifting the reference
wavelength by fixed amounts from one exposure to the next, the
selected time frame in our observations corresponds to 120~\AA. The
optimal spectral range including the highest number of lines and
ions in the widest temperature range within 120~\AA\ for detector~B
is the shortest wavelength range of 670-790~\AA. To further sample
the corona, we also included the two \ion[Fe xii] lines
1241.95~\AA~and 1349.43~\AA, even though they are far apart from the
selected interval. This choice makes some sense since coronal lines
might experience less temporal variation than the TR ones, so they
might still be representative of the corona.

\begin{figure*}
\centering
\includegraphics[width=0.35\textwidth,angle=90]{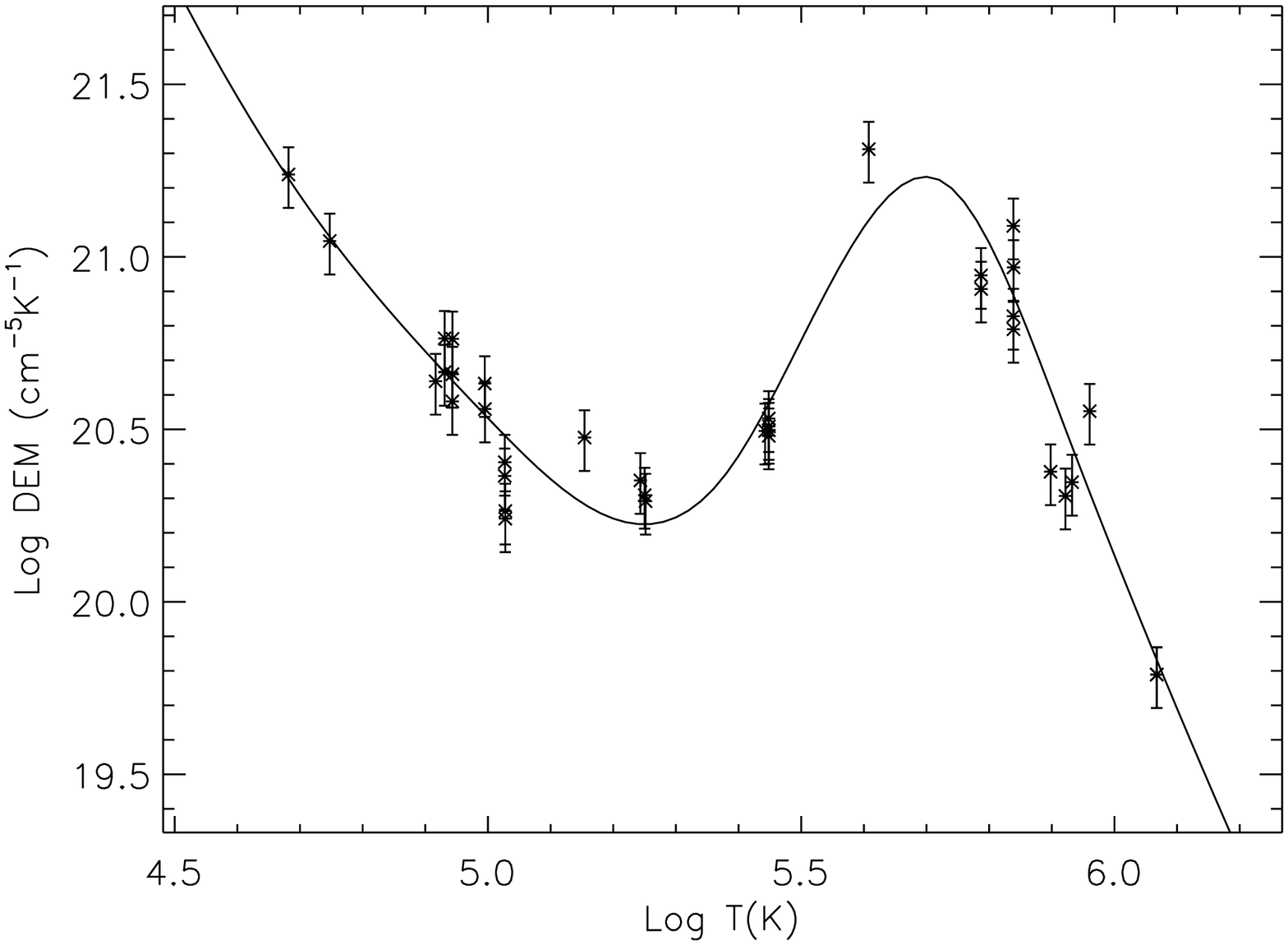}
\includegraphics[width=0.35\textwidth,angle=90]{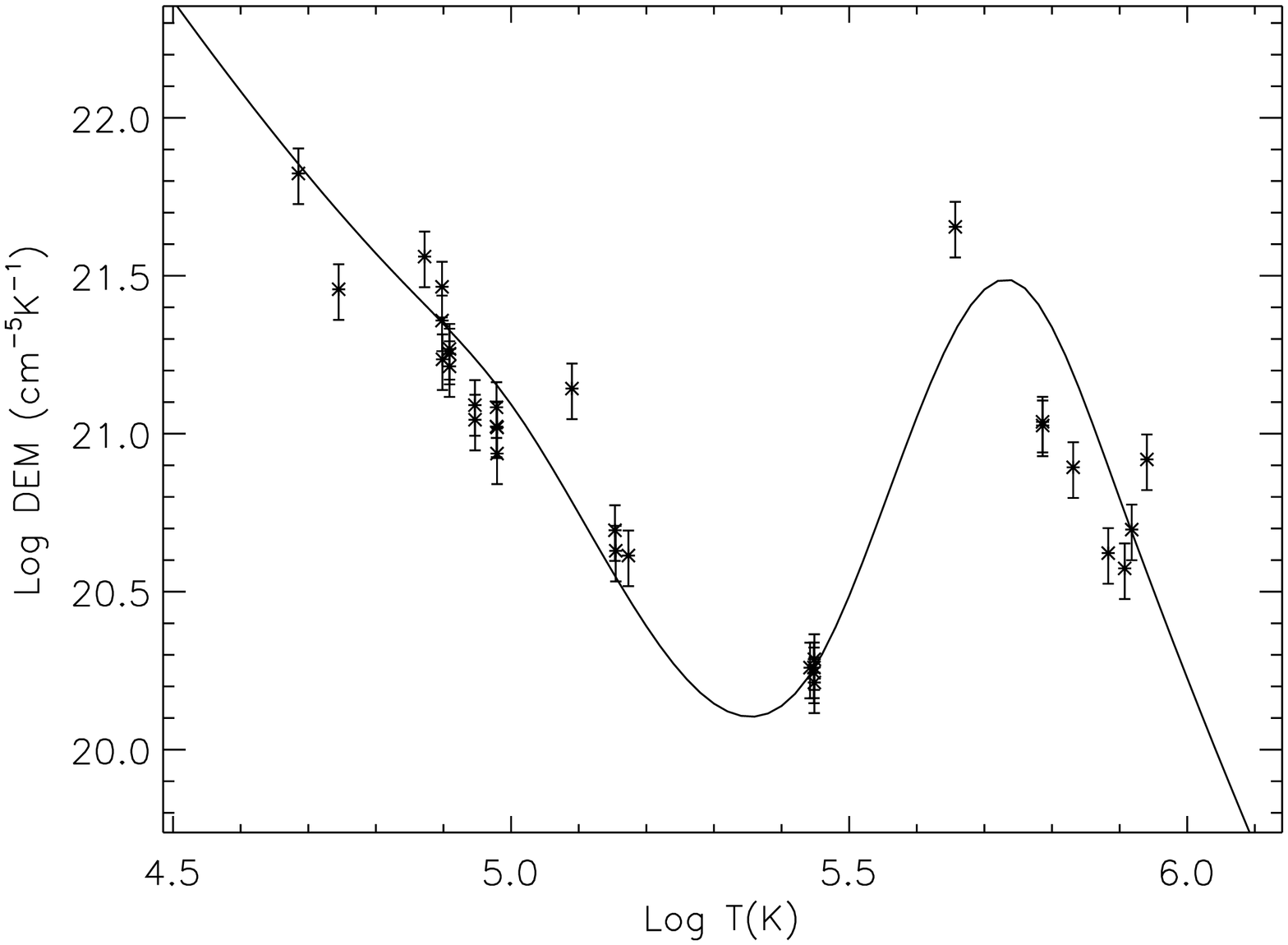}
\includegraphics[width=0.35\textwidth,angle=90]{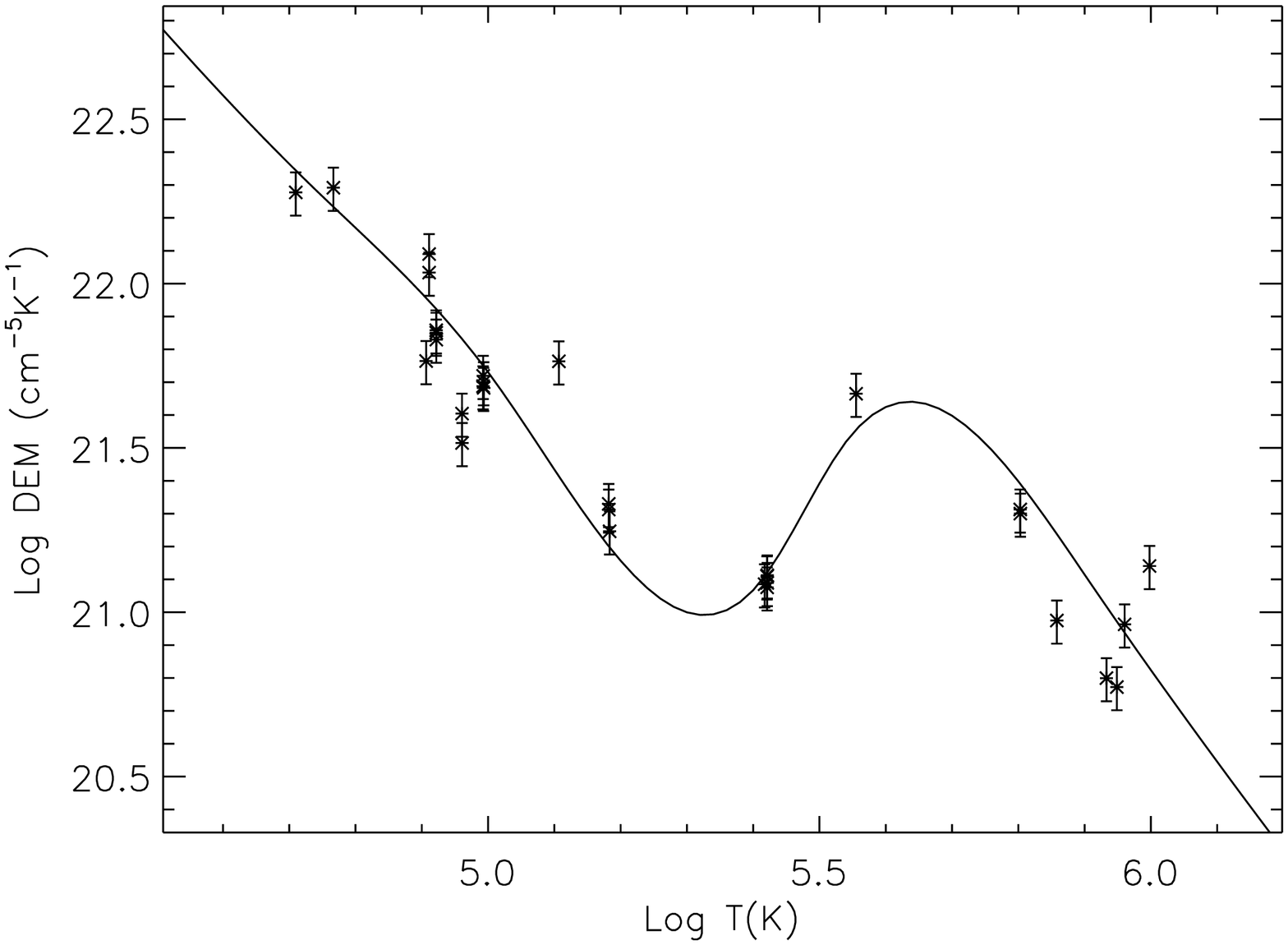}
\includegraphics[width=0.35\textwidth,angle=90]{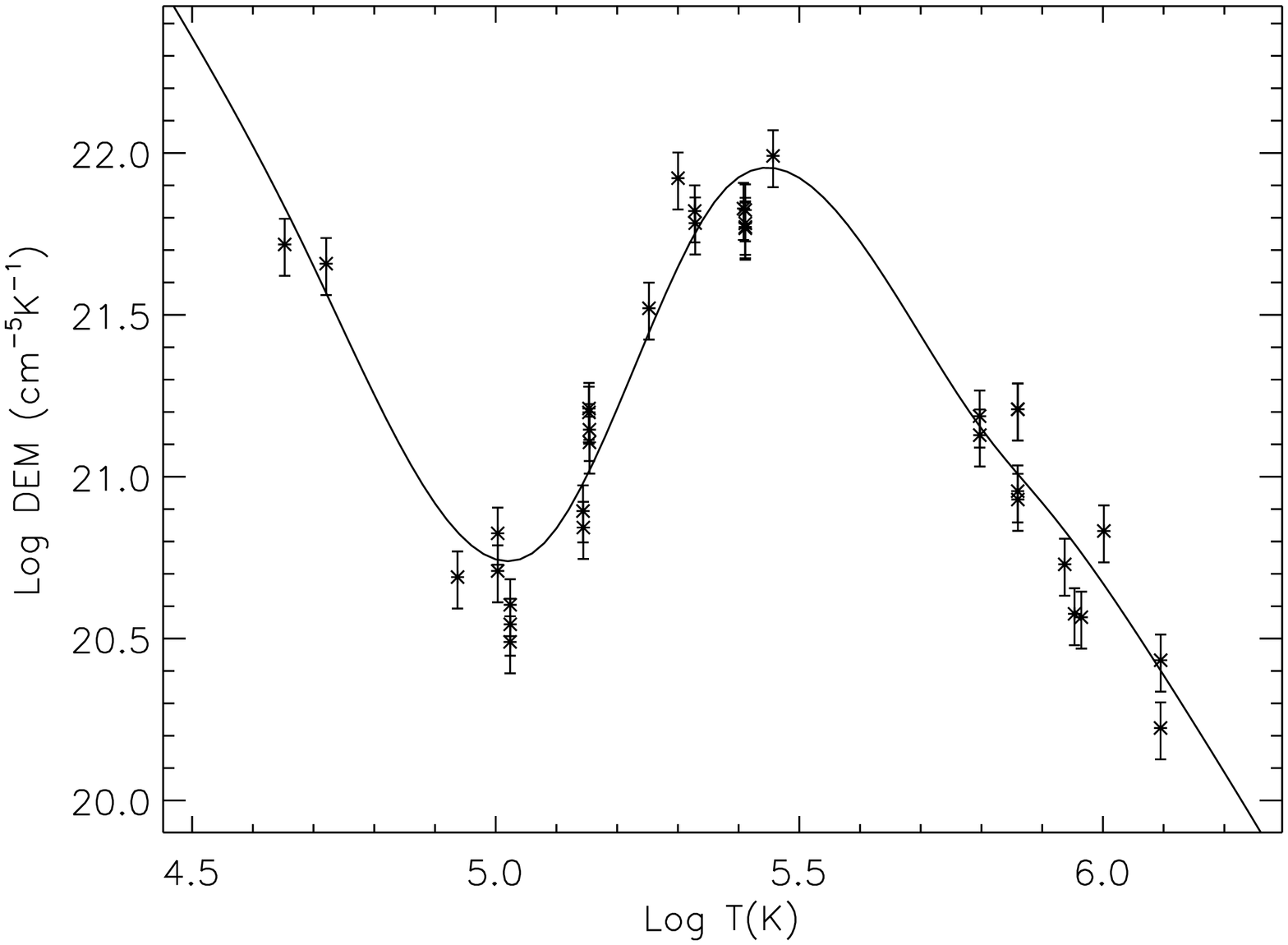}
\caption{~DEM curves for the four regions. {\bf Top left:} Umbra;
{\bf Top right:} Penumbra; {\bf Bottom left}: Plage; {\bf Bottom
right:} Plume.}\label{fig4}
\end{figure*}

We have applied the DEM diagnostic technique to the lines observed
between 670~\AA\ and 790~\AA\ in first order, and this time the
solutions converged to the curves displayed in Fig.~\ref{fig4}.
However, there was one last correction necessary to determine the
final DEM curves: the abundances of the low-FIP elements (those with
a First Ionization Potential lower than 10~eV) needed to be
increased by a factor of 10. This huge factor is 2.5 times higher
than the the factor required by the normal FIP effect. This
correction only affects the corona, since below $\log
(T/\rm{K})=5.7$ all the ions belong to high-FIP elements; in
contrast, all coronal ions except \ion[Ne viii] belong to low-FIP
elements. This correction was required to ensure agreement between
results derived for the \ion[Mg viii] and \ion[Ne viii] lines, whose
temperatures of formation are very similar. It is important to note
that \ion[Ne viii] belongs to the lithium isoelectronic sequence.
This sequence has been found to overestimate theoretical
emissivities relative to those of the other elements, leading to EM
and DEM measurements lower by a factor $\simeq 2$ than those of
elements formed at similar temperatures \citep{Dupree72,Landi02}. If
we take this systematic effect into account for \ion[Ne viii], the
effective increase in the abundance of low-FIP elements is a factor
of 5, much closer to the standard FIP effect commonly measured in
the corona (i.e., \citet{Feldman00}).

All the DEM curves that we measured exhibited an enhanced peak at
$\log(T/\rm{K})\simeq 5.6-5.8$, in the upper transition region. The
only exception is the plume DEM, which peaks at lower temperatures,
$\log(T/\rm{K})=5.45$. The peak of the plume DEM curve exceeds the
DEM of other regions by one to two orders of magnitude at these
temperatures. All the curves are able to reproduce the observed
lines, but some scatter is still present among the measurements
provided by each line. It is difficult to say whether the DEM curves
have a coronal component beyond $\log(T/\rm{K})=6.0$, because the
restricted dataset of lines we used does not include real coronal
lines formed at temperatures higher than one million degrees. The
only exceptions are the two \ion[Fe xii] lines, but their
contribution function has a low temperature tail that is strongly
influenced by the colder plasma. Its emission measure is so large,
that the $\log T_{\rm{eff}}$ of the \ion[Fe xii] lines is lower than
the temperature of maximum abundance for \ion[Fe xii]. The curves
that we determined are compared with each other in Fig.~\ref{fig5}.
In absolute units (upper panel of Fig.~\ref{fig5}), the curves are
rather similar, the only exception being the plume curve, whose peak
is at lower temperatures. When normalized to their value at
$\log(T/\rm{K})=4.7$, the slope of all four curves is approximately
the same up to $\log(T/\rm{K})=5.0$. This behavior was noted by
\citet{Feldman09}. If we approximate the DEM curves below
$\log(T/\rm{K})=5.0$ as

\begin{equation}
DEM=a\log T + b
\end{equation}

\noindent the slope $a$ of the DEM is similar to the values found by
\citet{Feldman09} in coronal hole, quiet Sun, and active region
spectra. We note that in \citet{BrosiusLandi05} the DEM at lower
temperatures (less than $\log(T/\rm{K})=5.0$) was poorly determined
because of the lack of low-temperature lines. Our calculation
includes some lines within this temperature range and thus can
determine accurately the corresponding thermal structures. At higher
temperatures the normalized curves of umbra, penumbra, and plage are
all similar but the relative heights of their peaks are different,
the umbra DEM peak being higher than both the penumbra and the plage
one. The plume peak is the highest of all, although it occurs at
lower temperatures.

The plume DEM curve that we measured and the one determined by
\citet{BrosiusLandi05} are compared in Fig.~\ref{fig6}.
\citet{BrosiusLandi05} measured the DEM of a plume in two different
days, and found similar results: their DEM curves were broad and
stretched in the entire $5.4 \leq \log(T/\rm{K}) \leq 6.0$
temperature range. They also measured the DEM of the quiet Sun in
both days, and those curves, with a very narrow peak in the
$\log(T/\rm{K})= 6.1-6.3$ range, are also shown in Fig.~\ref{fig6}
for comparison. The plume DEM we measured (solid line in
Fig.~\ref{fig6}) has a much narrower peak at slightly lower
temperatures. One reason for these differences can of course be an
intrinsic variability in the sunspot plume thermal structure.
Unfortunately, to the best of our knowledge no other DEM
measurements of plumes have been made, so this conclusion needs to
be confirmed. Another cause of discrepancy might be that
\citet{BrosiusLandi05} used CDS spectra to determine the curves in
Fig.~\ref{fig6}. In these spectra, many more lines formed at
$\log(T/\rm{K}) > 5.9$ were present, so that the coronal component
of the DEM was better constrained than in the present work. The
presence of a larger coronal component than in our plume curve might
alter the shape of the DEM at temperatures higher than
$\log(T/\rm{K})=5.8$. In general, we can say that the main signature
of sunspot plumes is a greatly enhanced DEM at transition region
temperatures, while at $\log(T/\rm{K}) < 5.0$ the slope of the DEM
curve is similar to that of any other region in the Sun, indicating
that the thermal structures across this temperature range are fairly
similar.

\begin{figure}[tp]
\centering
\begin{minipage}[t]{0.48\textwidth}
\resizebox{\hsize}{!}
{\includegraphics[width=\textwidth,angle=90]{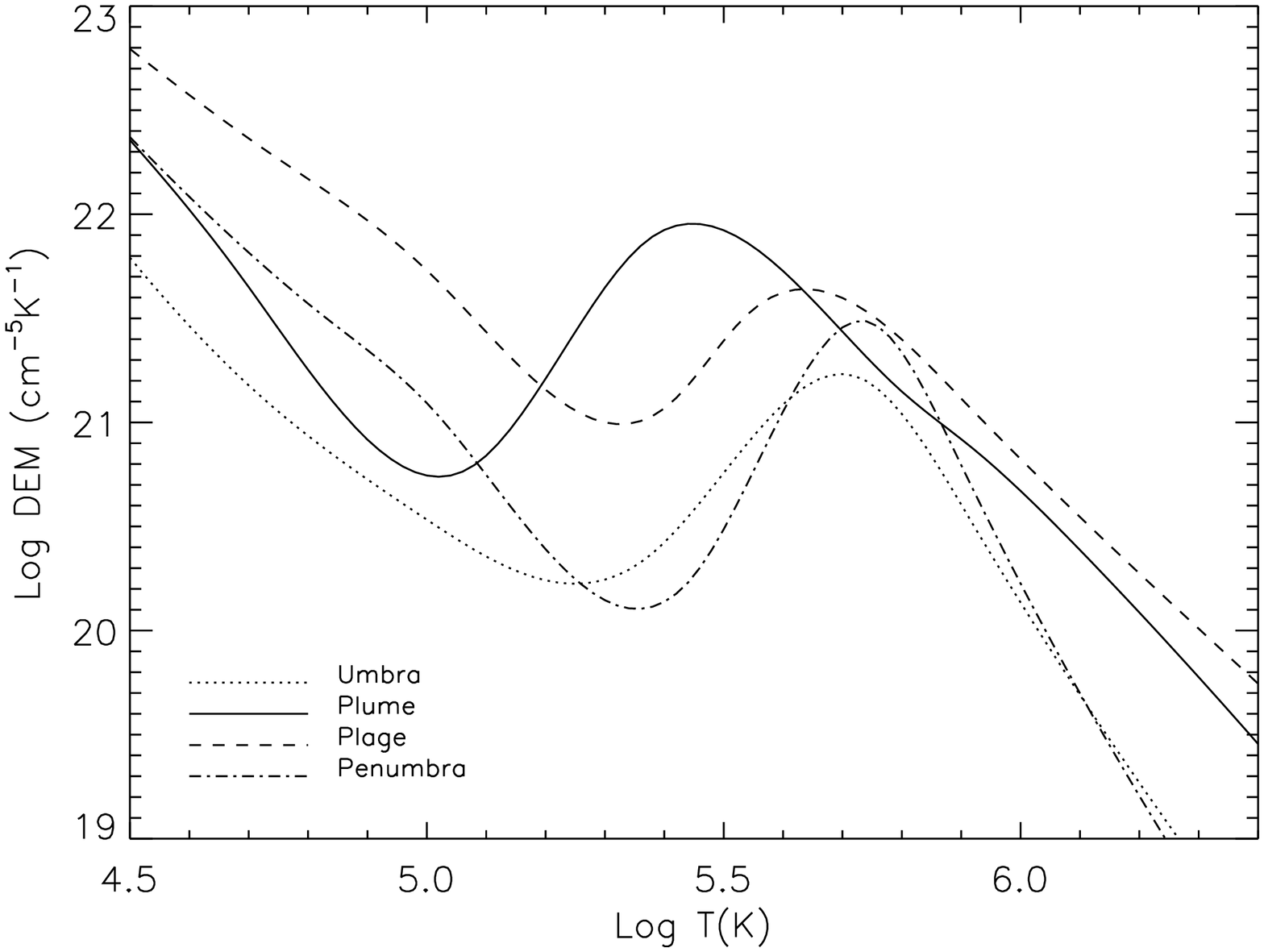}}
\end{minipage}
\begin{minipage}[t]{0.48\textwidth}
\resizebox{\hsize}{!}
{\includegraphics[width=\textwidth,angle=90]{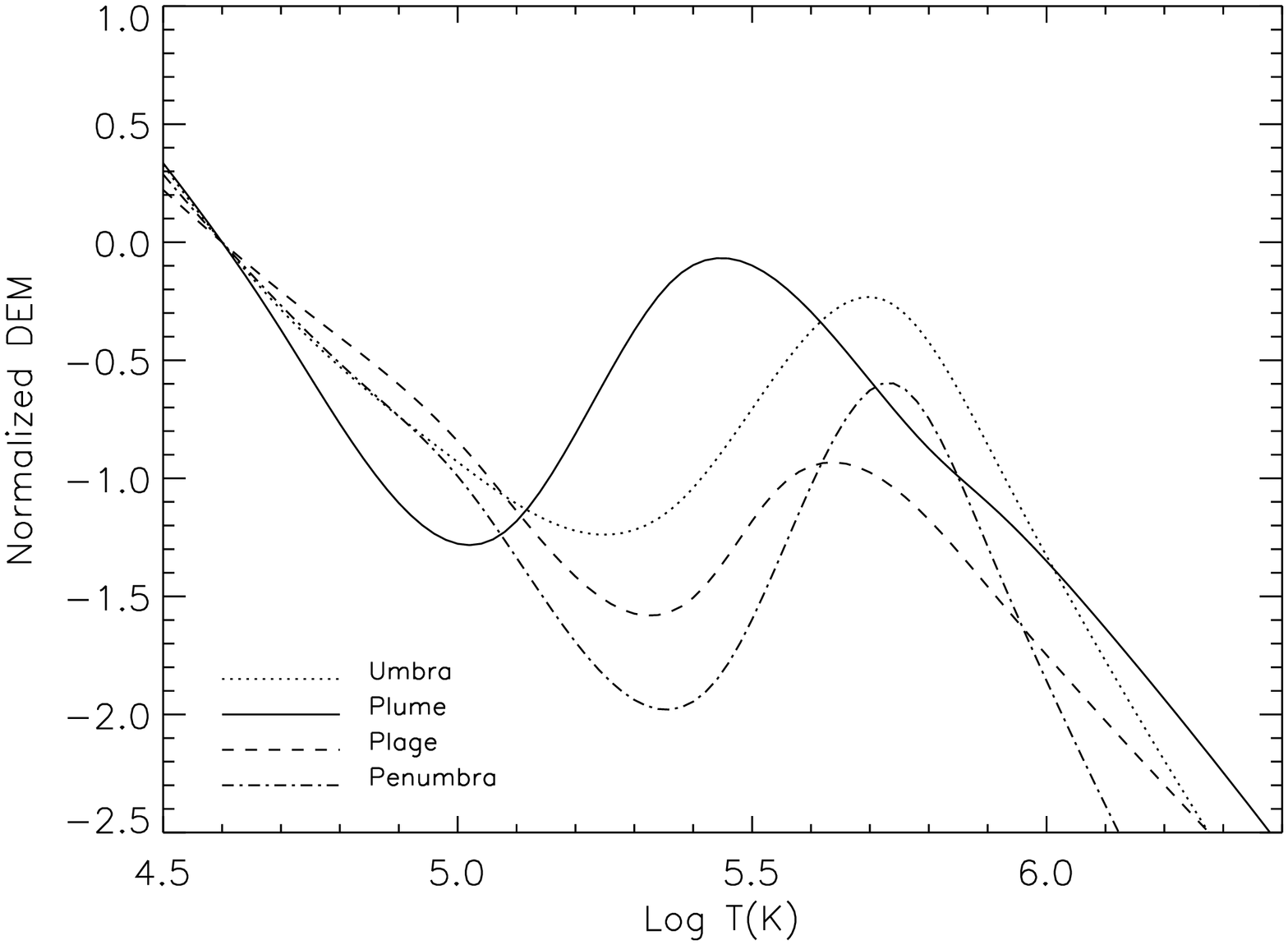}}
\end{minipage}
\begin{minipage}[b]{0.48\textwidth}
\caption{~Comparison of the DEM curves from the four regions. {\bf
Upper:} Absolute values; {\bf Lower:} Normalized values.}
\label{fig5}
\end{minipage}
\end{figure}

\begin{figure}
\resizebox{\hsize}{!}{\includegraphics[angle=90]{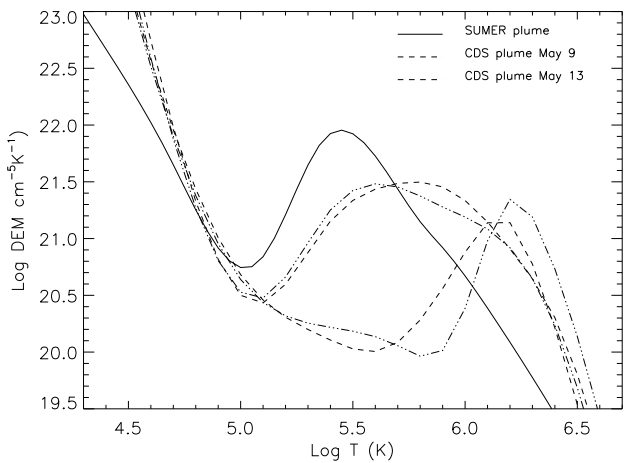}}
\caption{Comparison of the plume DEM curve with the results of
\citet{BrosiusLandi05}.} \label{fig6}
\end{figure}

We have to mention that the density and DEM calculations are based
on the assumption of ionization equilibrium. In the TR with the
strong temperature gradient and presence of significant flows, this
assumption is not necessarily valid \citep{Wilhelm04}. This, as well
as the inhomogeneity and variability of the TR plasma, might be
important to the determination of electron densities and DEM.
However, recent investigations using 3-D model indicated that the
assumption of ionization equilibrium might not be too bad in the
coronal and TR plasma \citep{Peter04,Peter06}. The calculations in
\cite{Peter06} showed that the ionization and recombination times
are at least not (much) longer and often shorter than the typical
hydrodynamic timescales. \cite{Peter06} also found that flows with
typical values of 10~km/s lead to a more shallow temperature
gradient and thus help to maintain an ionization equilibrium. Thus,
the approximation of ionization equilibrium in our study should not
be too bad.

\section{Filling factor}

Although it is well known that the emission of sunspot plumes is
greatly enhanced at TR temperatures, the reason for the enhanced
intensity remains unknown.

Our measurement reveals a higher intensity in the plume than in the
sunspot umbra, penumbra, and the surrounding plage region. The
enhanced intensity cannot result from a higher density, since the
derived TR densities of the plume are similar to those of the umbra
and even much lower than those of the penumbra and plume. It might
therefore be an effect of filling factor or temperature, or both.

By using temperature-sensitive line pairs, \citet{Doyle85} derived
the electron temperatures in sunspot plumes. They concluded that the
ions are shifted to lower temperatures in the sunspot plume. In our
reference spectra, we have no suitable temperature-sensitive line
pairs that were observed simultaneously. Thus, this conclusion
cannot be examined. However, our DEM analysis clearly reveals that
the emitting material of the sunspot plume is concentrated at lower
temperatures (higher peak at low \emph{T}), which is indicative of a
lower plume temperature, compared to that of the surrounding plasma.

With the measurements of line intensities and electron densities, we
are able to calculate the filling factor. As mentioned in
\citet{Dere08}, the observed intensity of a spectral line emitted by
a coronal loop is

\begin{equation}
\emph{\large{$I=\frac{1}{4\pi}\int~G(T,N_{\rm{e}})N_{\rm{e}}N_{\rm{H}}dh$}}
\label{equation1},
\end{equation}

\noindent where $G(T)$, $N_{\rm{e}}$, $N_{\rm{H}}$, and $dh$
represent the contribution function, electron density, hydrogen
density, and the differential of the integration, respectively.
Based on the assumption of an isothermal emission feature, we can
evaluate the contribution function at its peak temperature
$T_{\rm{max}}$. The ratio of the hydrogen density relative to the
electron density is about 0.8 in a completely ionized plasma of
cosmic composition \citep{Landi97}. Usually a loop is not completely
filled with emitting plasma, thus a volumetric filling factor should
be introduced to account for this. Taking into account these
considerations, Eq~(\ref{equation1}) can be rewritten as

\begin{equation}
\emph{\large{$I=\frac{0.8}{4\pi}G(T_{\rm{max}},N_{\rm{e}})N_{\rm{e}}^{2}fL$}}
\label{equation2},
\end{equation}

\noindent where $f$ is the filling factor and $L$ is the length of
the integration path.

We used four O~{\sc{v}} lines to determine the filling factor of the
sunspot plume. The O~{\sc{v}}~774.51~{\AA} line is density
independent, and the other three lines have been used to measure the
electron densities. Table~\ref{table3} lists the calculation details
from Eq~(\ref{equation2}). Here the intensity refers to that
averaged over the plume part along the slit and was converted into
energy units by using the procedure of \emph{radiometry.pro} in
\emph{SSW} (SolarSoft). The contribution function at its peak
temperature $T_{\rm{max}}$ is density dependent and was calculated
at the corresponding density from the CHIANTI database. The factor
$1/{4\pi}$ is included in the $G(T_{\rm{max}},N_{\rm{e}})$
calculation.

It is difficult to estimate $L$. However, we may be able to provide
an upper and a lower limit to this parameter. In the case of
ionization equilibrium, the major part of the O~{\sc{v}} emission
comes from plasmas in the range of 5.2$~\leq\log(T/\rm{K})\leq~$5.5.
By using a new model of quiet-Sun chromosphere and transition
region, \citet{Avrett08} determined the distribution of temperature
with height. Using their results, the span of the plasma in the
range of 5.2$~\leq\log(T/\rm{K})\leq~$5.5 is around 500~km. We
assumed this value to be the lower limit to $L$. Since plumes might
be related to the high-arching loops observed in ARs (active
regions), we may use the extension of the O~{\sc{v}} emission in
cross-limb AR loop observations as the upper limit to $L$. An
example of this type of observation can be found in
\citet{Brekke97}, where the AR loop width is about
10$^{\prime\prime}$. Thus, we adopted a value of 7~Mm as the upper
limit.

\begin {table}[]
\caption[]{The details of the filling factor calculations, for the
O~{\sc{v}} emission of the sunspot plume observed in 2006. The units
of $I$, $N_{\rm{e}}$ and $G(T_{\rm{max}},N_{\rm{e}})$ are
$\rm{erg~cm^{-2}~s^{-1}~sr^{-1}}$, $\rm{cm}^{-3}$ and
$\rm{erg~cm^{-3}~s^{-1}~sr^{-1}}$, respectively. } \label{table3}
\centering
\begin {tabular}{cccrcc}
\hline\hline
line (${\AA}$) & $I$ & $\rm{log}$~$N_{\rm{e}}$ & $G(T_{\rm{max}},N_{\rm{e}})$  \vline& $L~\rm{(Mm)}$ & $f$\\
\hline
758.68 & 190.77 & 10.12 & $2.81\times10^{-25}$  \vline& 0.5 & 0.096\\
 &  &  &   \vline& 7.0 & 0.007\\
\hline
759.43 & 154.72 & 10.09 & $2.15\times10^{-25}$  \vline& 0.5 & 0.120\\
 &  &  &   \vline& 7.0 & 0.009\\
\hline
761.99 & 202.43 & 10.03 & $2.61\times10^{-25}$  \vline& 0.5 & 0.160\\
 &  &  &   \vline& 7.0 & 0.012\\
\hline
774.51 & 44.19  & 10.03 & $5.54\times10^{-26}$  \vline& 0.5 & 0.170\\
 &  &  &   \vline& 7.0 & 0.012\\
\hline
\end {tabular}
\end {table}

The derived filling factors turn out to be relatively large, which
might explain the very strong TR emission of sunspot plumes. To
compare the filling factor of the plume with those of other
features, we can assume that the integration paths in various
regions are the same, although this assumption might not be
realistic due to a possible difference in their thermal and magnetic
structures. Again we obtained the average intensities of the four
regions, and calculated $G(T_{\rm{max}})$ at corresponding densities
by using the CHIANTI database. The ratios of the filling factors in
the four regions are listed in Table~\ref{table4}.

\begin {table}[]
\caption[]{The ratios of the filling factors in the four regions,
for the O~{\sc{v}} emission in the 2006 data set.} \label{table4}
\centering
\begin {tabular}{rc}
\hline\hline
line (${\AA}$) \vline& ratio (umbra : penumbra : plage : plume)\\
\hline
758.68  \vline& 70.8 : 1.0 : 6.3 : 846.2\\
\hline
759.43  \vline& 211.8 : 1.0 : 9.6 : 2058.8\\
\hline
761.99  \vline& 114.3 : 1.0 : 5.9 : 1507.9\\
\hline
774.51  \vline& 69.3 : 1.0 : 2.4 : 666.7\\
\hline
\end {tabular}
\end {table}

The results show that the filling factor of the sunspot plume at TR
temperatures is between one and three orders of magnitude higher
than for the surrounding regions. The difference is so large that it
seems safe to conclude that the strong TR emission of sunspot plumes
is mainly the result of a large filling factor. However, we should
not exclude the possibility that other effects, e.g., the region
emitting TR lines being much thicker in sunspot plumes, might also
play a role in producing the strong emission.

\section{Peculiar lines in the sunspot}

As mentioned in our earlier paper \citep{Curdt00}, more than 100
``peculiar" lines including several H$_{2}$ lines are present in the
sunspot reference spectra obtained on March 18, 1999. These lines
are also found in the sunspot atlas presented by \citet{Curdt01}.
Most of these lines are upper-TR lines with a temperature range of
5.3$~\leq(\log(T/\rm{K})\leq~$6.0, corresponding to 4 to 8-fold
ionized species. Here we confirm that these ``peculiar" lines are
also present in all of the 5 analyzed data sets. These lines are
weak or not observed in either the quiet Sun and or corona. Some of
them may be present in streamer spectra.

Many of these ``peculiar" lines correspond to forbidden transitions.
Thus, we may speculate that an extremely low density might be
responsible for their formation. However, as we mentioned
previously, the umbra and plume densities derived here are similar
to, or slightly higher than the density of the normal quiet Sun.
This means that the sunspot plasma density is not extremely low and
the above explanation is questionable.


Through a visual inspection of the spectra analyzed here and several
other sunspot reference spectra, we found that the ``peculiar" lines
seem more likely to be associated with plumes, rather than the
umbrae. One possible reason why these lines are present in the umbra
\citep{Curdt01} is that part of the sunspot plumes are just located
above the umbra and the sample of the sunspot emission might be a
mixture of the plume and umbra emissions. Since the plume emission
is much stronger than the umbra emission at TR temperatures, the
average sunspot spectrum in \citet{Curdt01} is dominated by the
plume spectrum.

The reason why these ``peculiar" lines are so distinctive in the
sunspot plume is probably a combination of two effects. First, the
plume emission is strongest at upper-TR temperatures. Second, the
continuum emission of the plume is much weaker than that of the
plage.

\begin{figure*}
\resizebox{\hsize}{!}{\includegraphics{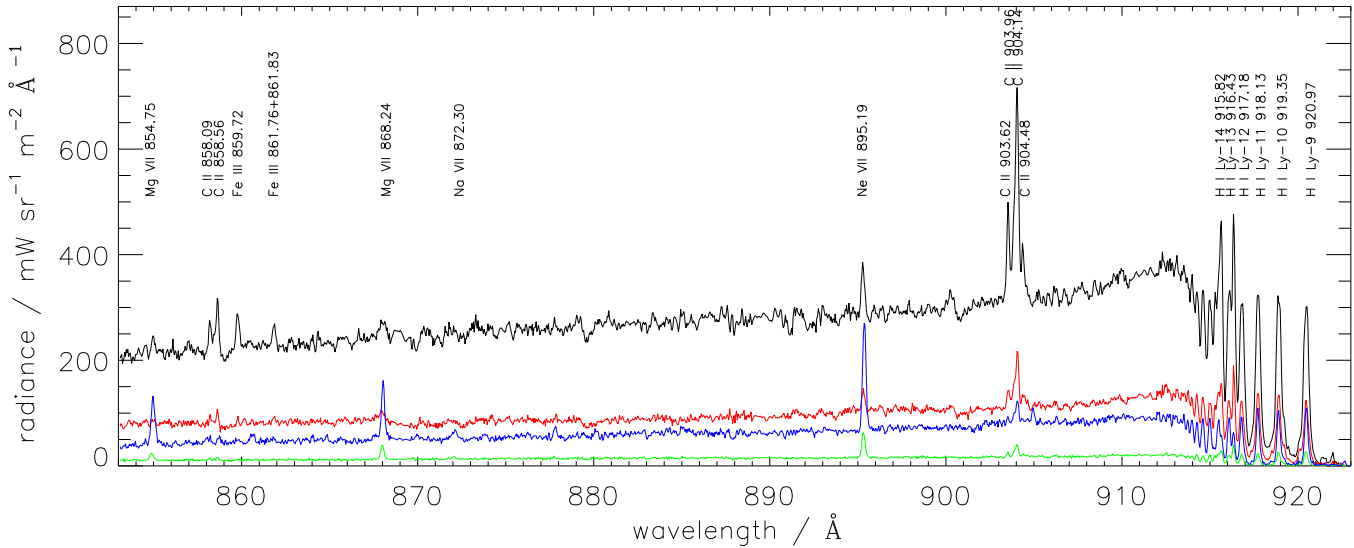}} \caption{Part
of the spectra observed in the plage, penumbra, umbra and plume for
the 2006 data set. The black, red, green and blue lines correspond
to the plage, penumbra, umbra and plume spectra.} \label{fig.7}
\end{figure*}

We list these ``peculiar" lines in Table 6. As an example, in
Fig.~\ref{fig.7} we show part of the spectra observed in the plage,
penumbra, umbra, and plume for the 2006 data set. We see that
several forbidden lines (e.g., Mg~{\sc{vii}}~854.75~{\AA},
Mg~{\sc{vii}}~868.24~{\AA}, Ne~{\sc{vii}}~895.19~{\AA}) become very
prominent in the plume spectrum. Here, we also see in the umbra
spectrum the enhanced forbidden lines, which are not so enhanced as
in the plume and might be due to contamination from the plume
emission at the formation temperatures of these lines. However, the
strongly reduced continuum emission of the umbra may also cause the
forbidden lines to be distinctive.

The H$_{2}$ emission is extremely weak in the quiet Sun
\citep{Sandlin86}. It was also observed in flares \citep{Bartoe79}
and microflares in the active region plage \citep{Innes08}. In
sunspot regions, the H$_{2}$ emission is distinctive and relatively
strong \citep{Jordan78,Schueler1999,Curdt01}. Several lines of the
H$_{2}$ Werner bands are within the SUMER spectral range. These
lines are believed to be excited by resonance fluorescence through
the strong O~{\sc{vi}}~1031.9~{\AA} line. In our data, the emission
of these lines is clearly detected in the umbra and plume regions,
which might be related to the reduced opacity above the sunspot. The
O~{\sc{vi}}~1031.9~{\AA} emission is able to reach the chromosphere
and excite the H$_{2}$ lines if the opacity is low.

\section{Summary}

By analyzing sunspot reference spectra taken by the SUMER
instrument, we have shown that the TR above sunspots has some
distinctive properties compared to the surrounding plage regions.

We have found that the hydrogen Lyman line profiles are not reversed
in sunspots at different locations (heliocentric angle up to
$49^\circ$) on the solar disk. The Lyman lines are also not reversed
in sunspot plumes. In the plage region, the lower Lyman line
profiles are obviously reversed, a phenomenon also found in the
normal quiet Sun. Line-pair diagnostics yields an electron density
of $\log(N_{\rm{e}}/\rm{cm}^{-3})\approx10.0$ for the umbra and the
plume, and $\approx11.0$ for the penumbra and plage, at TR
temperatures. To explain these results, we suggest that the TR above
sunspots is higher and probably more extended, and the opacity above
sunspots is much lower than in the TR above plage regions.

We also completed a DEM analysis for the sunspot observed in 2006.
To the best of our knowledge, this is the first time that SUMER
spectra have been used for the DEM diagnostics of a sunspot. The DEM
curve of the plume is obviously different from those of other
regions. It peaks at a lower temperature of around
$\log(T/\rm{K})=5.45$, which exceeds the DEM of other regions by one
to two orders of magnitude at these temperatures. At $\log(T/\rm{K})
< 5.0$, the slope of the DEM curve is similar in the four regions,
indicating that the thermal structure in this temperature range is
fairly similar everywhere in and around the sunspots.

The reason why the plume emission is so strong at upper-TR
temperatures has been investigated for the first time. Our
calculations seem to indicate that the enhanced TR emission of the
sunspot plume is very likely to be the result of a large filling
factor.

More than 100 lines that are rather weak or not observed anywhere
else on the Sun, are well observed by SUMER in the sunspot,
especially in the sunspot plume. We propose that it is the
combination of strongly enhanced emission at TR temperatures and
reduced continuum cause these normally weak lines to be clearly
distinctive in the spectra of sunspot plumes.

\begin{acknowledgements}
The SUMER project is financially supported by DLR, CNES, NASA, and
the ESA PRODEX Programme (Swiss contribution). SUMER is an
instrument onboard {\it SOHO}, a mission operated by ESA and NASA.
We thank Dr. D. E. Innes and Dr. H. Peter for the helpful comments.

Hui Tian is supported by the IMPRS graduate school run jointly by
the Max Planck Society and the Universities of G\"ottingen and
Braunschweig. The work of Hui Tian's group at PKU is supported by
the National Natural Science Foundation of China (NSFC) under
contract 40874090. Enrico Landi acknowledges support from the
NNG06EA14I, NNH06CD24C NASA grants.

\end{acknowledgements}

\end{document}


\begin{table*}[]
\caption[]{} \label{table1} \centering
\begin{tabular}{lrcllll}
\end{tabular}
\end{table*}

\begin{table*}[]
\caption[]{} \label{table2} \centering
\begin{tabular}{lrcllll}
\end{tabular}
\end{table*}

\begin{table*}[]
\caption[]{} \label{table3} \centering
\begin{tabular}{lrcllll}
\end{tabular}
\end{table*}

\begin{table*}[]
\caption[]{} \label{table4} \centering
\begin{tabular}{lrcllll}
\end{tabular}
\end{table*}

\begin{deluxetable}{cccccccc}
\tablewidth{0pt} \tabletypesize{\scriptsize} \tablecaption{~List of
lines used for DEM diagnostics. The values of intensities (in the
unit of $\rm{photon~sr^{-1} s^{-1} arcsec^{-2}}$) of the four
regions are also shown. \label{table5}} \tablehead{\colhead{Ion} &
\colhead{$\lambda_{obs}$} & \colhead{Transition} &
\colhead{$\log(T_f/\rm{K}$)} & \colhead{Umbra} & \colhead{Penumbra}
& \colhead{Plage} & \colhead{Plume} } \tablecolumns{8} \startdata
     \ion[C ii]  &    903.590 & \orb[2s 2]\orb[2p ] \tm[2 P 1/2] - \orb[2s2p 2] \tm[2 P 3/2] & 4.37 & 1.924e+00 &  1.018e+01 &  4.125e+01 &  4.625e+00 \\
     \ion[C ii]  &    903.990 & \orb[2s 2]\orb[2p ] \tm[2 P 1/2] - \orb[2s2p 2] \tm[2 P 1/2] &  & 1.826e+00 &  1.124e+01 &  2.636e+01 &  3.832e+00 \\
     \ion[C ii]  &    904.140 & \orb[2s 2]\orb[2p ] \tm[2 P 3/2] - \orb[2s2p 2] \tm[2 P 3/2] &  & 5.342e+00 &  2.528e+01 &  8.981e+01 &  1.010e+01 \\
     \ion[C ii]  &    904.460 & \orb[2s 2]\orb[2p ] \tm[2 P 3/2] - \orb[2s2p 2] \tm[2 P 1/2] &  & 1.106e+00 &  9.160e+00 &  2.566e+01 &  3.981e+00 \\
     \ion[C ii]  &   1036.340 & \orb[2s 2]\orb[2p ] \tm[2 P 1/2] - \orb[2s2p 2] \tm[2 S 1/2] &  & 2.527e+01 &  9.625e+01 &  4.045e+02 &  3.452e+01 \\
     \ion[C ii]  &   1037.000 & \orb[2s 2]\orb[2p ] \tm[2 P 3/2] - \orb[2s2p 2] \tm[2 S 1/2] &  & 3.181e+01 &  1.429e+02 &  4.705e+02 &  4.227e+01 \\
     \ion[C ii]  &   1334.530 & \orb[2s 2]\orb[2p ] \tm[2 P 1/2] - \orb[2s2p 2] \tm[2 D 3/2] &  & 7.287e+02 &  8.435e+03 &  2.665e+04 &  1.375e+03 \\
     \ion[C ii]  &   1335.710 & \orb[2s 2]\orb[2p ] \tm[2 P 3/2] - \orb[2s2p 2] \tm[2 D 5/2] &  & 1.187e+03 &  1.161e+04 &  3.852e+04 &  2.237e+03 \\
    \ion[C iii]  &    977.030 & \orb[2s 2] \tm[1 S 0] - \orb[2s2p ] \tm[1 P 1] & 4.84 & 5.350e+02 &  2.381e+03 &  9.698e+03 &  6.622e+02 \\
    \ion[C iii]  &   1174.880 & \orb[2s2p ] \tm[3 P 1] - \orb[2p 2] \tm[3 P 2] &  & 3.734e+01 &  1.693e+02 &  9.485e+02 &  4.486e+01 \\
    \ion[C iii]  &   1175.240 & \orb[2s2p ] \tm[3 P 0] - \orb[2p 2] \tm[3 P 1] &  & 3.090e+01 &  1.371e+02 &  8.754e+02 &  3.823e+01 \\
    \ion[C iii]  &   1175.980 & \orb[2s2p ] \tm[3 P 1] - \orb[2p 2] \tm[3 P 0] &  & 2.954e+01 &  1.301e+02 &  8.114e+02 &  3.805e+01 \\
    \ion[C iii]  &   1176.370 & \orb[2s2p ] \tm[3 P 2] - \orb[2p 2] \tm[3 P 1] &  & 3.493e+01 &  1.690e+02 &  8.597e+02 &  4.494e+01 \\
    \ion[C iii]  &    690.530 & \orb[2s2p ] \tm[1 P 1] - \orb[2p 2] \tm[1 S 0] &  & 8.225e-01 &  2.587e+00 &  9.027e+00 &  1.404e+00 \\
     \ion[N ii]  &   1083.990 & \orb[2s 2]\orb[2p 2] \tm[3 P 0] - \orb[2s2p 3] \tm[3 D 1] & 4.43 & 1.064e+01 &  5.329e+01 &  1.531e+02 &  1.347e+01 \\
     \ion[N ii]  &   1085.540 & \orb[2s 2]\orb[2p 2] \tm[3 P 1] - \orb[2s2p 3] \tm[3 D 1,2] &  & 9.890e+00 &  6.396e+01 &  2.255e+02 &  9.610e+00 \\
     \ion[N ii]  &   1085.710 & \orb[2s 2]\orb[2p 2] \tm[3 P 1] - \orb[2s2p 3] \tm[3 D 3] &  & 4.385e+01 &  1.719e+02 &  5.054e+02 &  5.968e+01 \\
     \ion[N ii]  &    746.990 & \orb[2s 2]\orb[2p 2] \tm[1 D 2] - \orb[2s 2]\orb[2p3s ] \tm[1 P 1] &  & 4.308e-01 &  1.559e+00 &  4.913e+00 &  9.306e-01 \\
    \ion[N iii]  &    684.990 & \orb[2s 2]\orb[2p ] \tm[2 P 1/2] - \orb[2s2p 2] \tm[2 P 3/2] & 4.88 & 1.970e+00 &  4.252e+00 &  1.770e+01 &  3.062e+00 \\
    \ion[N iii]  &    685.500 & \orb[2s 2]\orb[2p ] \tm[2 P 1/2] - \orb[2s2p 2] \tm[2 P 1/2] &  & 2.547e+00 &  7.454e+00 &  3.300e+01 &  4.621e+00 \\
    \ion[N iii]  &    685.790 & \orb[2s 2]\orb[2p ] \tm[2 P 3/2] - \orb[2s2p 2] \tm[2 P 3/2] &  & 7.789e+00 &  2.054e+01 &  8.704e+01 &  1.334e+01 \\
    \ion[N iii]  &    763.330 & \orb[2s 2]\orb[2p ] \tm[2 P 1/2] - \orb[2s2p 2] \tm[2 S 1/2] &  & 1.822e+00 &  6.381e+00 &  2.865e+01 &  4.770e+00 \\
    \ion[N iii]  &    764.360 & \orb[2s 2]\orb[2p ] \tm[2 P 3/2] - \orb[2s2p 2] \tm[2 S 1/2] &  & 2.803e+00 &  9.623e+00 &  4.852e+01 &  7.042e+00 \\
    \ion[N iii]  &    989.820 & \orb[2s 2]\orb[2p ] \tm[2 P 1/2] - \orb[2s2p 2] \tm[2 D 3/2] &  & 1.345e+01 &  4.559e+01 &  1.764e+02 &  2.733e+01 \\
     \ion[N iv]  &    765.150 & \orb[2s 2] \tm[1 S 0] - \orb[2s2p ] \tm[1 P 1] & 5.15 & 5.743e+01 &  1.269e+02 &  6.247e+02 &  3.949e+02 \\
      \ion[N v]  &   1238.820 & \orb[2s ] \orb[2 S 1/2] - \orb[2p ] \tm[2 P 3/2] & 5.26 & 1.858e+02 &  5.114e+02 &  2.296e+03 &  1.093e+03 \\
      \ion[N v]  &   1242.800 & \orb[2s ] \orb[2 S 1/2] - \orb[2p ] \tm[2 P 1/2] &  & 8.740e+01 &  2.233e+02 &  1.046e+03 &  5.191e+02 \\
     \ion[O ii]  &    796.660 & \orb[2s 2]\orb[2p 3] \tm[2 P 1/2,3/2] - \orb[2s2p 4] \tm[2 D 3/2,5/2] & 4.45 & 1.531e+00 &  3.672e+00 &  2.705e+01 &  5.133e+00 \\
     \ion[O ii]  &    832.750 & \orb[2s 2]\orb[2p 3] \orb[4 S 3/2] - \orb[2s2p 4] \tm[4 P 1/2] &  & 3.860e+00 &  1.599e+01 &  5.618e+01 &  8.631e+00 \\
     \ion[O ii]  &    833.320 & \orb[2s 2]\orb[2p 3] \orb[4 S 3/2] - \orb[2s2p 4] \tm[4 P 3/2] &  & 1.024e+01 &  4.465e+01 &  1.791e+02 &  2.034e+01 \\
     \ion[O ii]  &    834.450 & \orb[2s 2]\orb[2p 3] \orb[4 S 3/2] - \orb[2s2p 4] \tm[4 P 5/2] &  & 1.349e+01 &  6.033e+01 &  2.460e+02 &  2.092e+01 \\
    \ion[O iii]  &    702.330 & \orb[2s 2]\orb[2p 2] \orb[3 P 0] - \orb[2s2p 3] \tm[3 P 1] & 4.93 & 6.518e+00 &  1.267e+01 &  4.592e+01 &  1.746e+01 \\
    \ion[O iii]  &    703.870 & \orb[2s 2]\orb[2p 2] \orb[3 P 1] - \orb[2s2p 3] \tm[3 P 1,2] &  & 2.763e+01 &  5.699e+01 &  1.870e+02 &  7.773e+01 \\
    \ion[O iii]  &    832.940 & \orb[2s 2]\orb[2p 2] \orb[3 P 0] - \orb[2s2p 3] \tm[3 D 1] &  & 7.761e+00 &  3.194e+01 &  1.661e+02 &  4.104e+01 \\
    \ion[O iii]  &    833.740 & \orb[2s 2]\orb[2p 2] \orb[3 P 1] - \orb[2s2p 3] \tm[3 D 1,2] &  & 1.880e+01 &  8.064e+01 &  3.955e+02 &  1.272e+02 \\
    \ion[O iii]  &    835.090 & \orb[2s 2]\orb[2p 2] \orb[3 P 2] - \orb[2s2p 3] \tm[3 D 1,2] &  & 6.567e+00 &  2.925e+01 &  8.284e+01 &  2.850e+01 \\
    \ion[O iii]  &    835.280 & \orb[2s 2]\orb[2p 2] \orb[3 P 2] - \orb[2s2p 3] \tm[3 D 3] &  & 2.757e+01 &  1.123e+02 &  6.207e+02 &  1.840e+02 \\
     \ion[O iv]  &    787.720 & \orb[2s 2]\orb[2p ] \tm[2 P 1/2] - \orb[2s2p 2] \tm[2 D 3/2] & 5.18 & 6.234e+01 &  8.764e+01 &  4.506e+02 &  9.172e+02 \\
     \ion[O iv]  &    790.190 & \orb[2s 2]\orb[2p ] \tm[2 P 3/2] - \orb[2s2p 2] \tm[2 D 3/2,5/2] &  & 1.291e+02 &  2.000e+02 &  1.032e+03 &  1.990e+03 \\
     \ion[O iv]  &   1397.220 & \orb[2s 2]\orb[2p ] \tm[2 P 1/2] - \orb[2s2p 2] \tm[4 P 3/2] &  & 4.660e+00 &  1.850e+01 &  2.898e+01 &  1.978e+01 \\
     \ion[O iv]  &   1399.770 & \orb[2s 2]\orb[2p ] \tm[2 P 1/2] - \orb[2s2p 2] \tm[4 P 1/2] &  & 1.378e+01 &  6.094e+01 &  1.703e+02 &  8.864e+01 \\
     \ion[O iv]  &   1401.160 & \orb[2s 2]\orb[2p ] \tm[2 P 3/2] - \orb[2s2p 2] \tm[4 P 5/2] &  & 6.494e+01 &  1.915e+02 &  5.799e+02 &  4.078e+02 \\
     \ion[O iv]  &   1407.390 & \orb[2s 2]\orb[2p ] \tm[2 P 3/2] - \orb[2s2p 2] \tm[4 P 1/2] &  & 1.629e+01 &  5.708e+01 &  2.056e+02 &  9.552e+01 \\
      \ion[O v]  &    758.680 & \orb[2s2p ] \tm[3 P 1] - \orb[2p 2] \tm[3 P 2] & 5.37 & 1.126e+01 &  1.565e+01 &  9.466e+01 &  1.829e+02 \\
      \ion[O v]  &    760.210 & \orb[2s2p ] \tm[3 P 1] - \orb[2p 2] \tm[3 P 1] &  & 6.276e+00 &  8.414e+00 &  5.152e+01 &  1.048e+02 \\
      \ion[O v]  &    760.430 & \orb[2s2p ] \tm[3 P 2] - \orb[2p 2] \tm[3 P 2] &  & 3.441e+01 &  3.935e+01 &  2.673e+02 &  5.590e+02 \\
      \ion[O v]  &    761.130 & \orb[2s2p ] \tm[3 P 1] - \orb[2p 2] \tm[3 P 0] &  & 1.958e+00 &  8.264e+00 &  4.919e+01 &  3.561e+01 \\
      \ion[O v]  &    761.990 & \orb[2s2p ] \tm[3 P 2] - \orb[2p 2] \tm[3 P 1] &  & 1.165e+01 &  1.446e+01 &  9.213e+01 &  1.973e+02 \\
     \ion[O vi]  &   1031.930 & \orb[2s ] \tm[2 S 1/2] - \orb[2p ] \tm[2 P 3/2] & 5.47 & 1.117e+03 &  1.080e+03 &  2.917e+03 &  1.097e+04 \\
     \ion[O vi]  &   1037.640 & \orb[2s ] \tm[2 S 1/2] - \orb[2p ] \tm[2 P 1/2] &  & 4.819e+02 &  4.710e+02 &  1.456e+03 &  4.477e+03 \\
   \ion[Ne vii]  &    895.170 & \orb[2s 2] \tm[1 S 0] - \orb[2s2p ] \tm[3 P 1] & 5.71 & 1.171e+01 &  1.513e+01 &  4.545e+01 &  6.226e+01 \\
  \ion[Ne viii]  &    770.420 & \orb[2s ] \tm[2 S 1/2] - \orb[2p ] \tm[2 P 3/2] & 5.80 & 1.202e+02 &  1.398e+02 &  3.742e+02 &  2.751e+02 \\
  \ion[Ne viii]  &    780.300 & \orb[2s ] \tm[2 S 1/2] - \orb[2p ] \tm[2 P 1/2] &  & 6.699e+01 &  7.310e+01 &  1.958e+02 &  1.600e+02 \\
    \ion[Na ix]  &    681.680 & \orb[2s ] \tm[2 S 1/2] - \orb[2p ] \tm[2 P 3/2] & 5.93 & 6.417e+00 &  9.360e+00 &  2.338e+01 &  1.956e+01 \\
     \ion[Mg v]  &   1324.590 & \orb[2s 2]\orb[2p 4] \tm[3 P 1] - \orb[2s 2]\orb[2p 4] \tm[1 S 0] & 5.43 & 5.103e+00 &            &            &  2.056e+01 \\
    \ion[Mg vi]  &   1191.680 & \orb[2s 2]\orb[2p 3] \tm[4 S 3/2] - \orb[2s 2]\orb[2p 3] \tm[2 P 1/2] & 5.64 & 5.339e+00 &            &            &  8.800e+00 \\
   \ion[Mg vii]  &    868.130 & \orb[2s 2]\orb[2p 2] \tm[3 P 2] - \orb[2s2p 3] \tm[5 S 2] & 5.80 & 6.372e+00 &  7.056e+00 &            &  4.244e+01 \\
  \ion[Mg viii]  &    762.650 & \orb[2s 2]\orb[2p ] \tm[2 P 1/2] - \orb[2s2p 2] \tm[4 P 3/2] & 5.90 & 9.264e-01 &  2.151e+00 &  4.827e+00 &  2.160e+00 \\
  \ion[Mg viii]  &    769.380 & \orb[2s 2]\orb[2p ] \tm[2 P 1/2] - \orb[2s2p 2] \tm[4 P 1/2] &  & 1.888e+00 &            &            &  3.403e+00 \\
  \ion[Mg viii]  &    782.340 & \orb[2s 2]\orb[2p ] \tm[2 P 3/2] - \orb[2s2p 2] \tm[4 P 3/2] &  & 5.775e+00 &  6.189e+00 &  1.091e+01 &  1.070e+01 \\
  \ion[Mg viii]  &    789.430 & \orb[2s 2]\orb[2p ] \tm[2 P 3/2] - \orb[2s2p 2] \tm[4 P 1/2] &  & 2.818e+00 &            &            &  4.980e+00 \\
    \ion[Mg ix]  &    706.020 & \orb[2s 2] \tm[1 S 0] - \orb[2s2p ] \tm[3 P 1] & 5.98 & 7.680e+00 &  1.195e+01 &  2.852e+01 &  1.813e+01 \\
    \ion[Mg ix]  &    749.540 & \orb[2s2p ] \tm[1 P 1] - \orb[2p 2] \tm[1 D 2] &  & 1.534e+00 &  2.893e+00 &  8.098e+00 &  3.214e+00 \\
   \ion[Si iii]  &   1206.510 & \orb[3s 2] \tm[1 S 0] - \orb[3s3p ] \tm[1 P 1] & 4.68 & 4.023e+02 &  2.882e+03 &  2.488e+04 &  1.853e+03 \\
   \ion[Si iii]  &   1294.580 & \orb[3s3p ] \tm[3 P 1] - \orb[3p 2] \tm[3 P 2] &  & 5.451e+00 &  4.369e+01 &  2.681e+02 &  1.926e+01 \\
   \ion[Si iii]  &   1296.770 & \orb[3s3p ] \tm[3 P 0] - \orb[3p 2] \tm[3 P 1] &  & 4.527e+00 &  3.720e+01 &  2.134e+02 &  1.712e+01 \\
   \ion[Si iii]  &   1298.960 & \orb[3s3p ] \tm[3 P 2] - \orb[3p 2] \tm[3 P 2] &  & 1.607e+01 &  1.208e+02 &  9.304e+02 &  7.092e+01 \\
   \ion[Si iii]  &   1301.160 & \orb[3s3p ] \tm[3 P 1] - \orb[3p 2] \tm[3 P 0] &  & 3.316e+00 &  3.012e+01 &  1.898e+02 &  1.488e+01 \\
    \ion[Si iv]  &    815.050 & \orb[3p ] \tm[2 P 1/2] - \orb[4s ] \tm[2 S 1/2] & 4.86 & 1.086e+00 &  4.641e+00 &  1.105e+01 &  3.598e+00 \\
    \ion[Si iv]  &    818.150 & \orb[3p ] \tm[2 P 3/2] - \orb[4s ] \tm[2 S 1/2] &  & 2.399e+00 &  1.048e+01 &  2.952e+01 &  9.645e+00 \\
    \ion[Si iv]  &   1128.350 & \orb[3p ] \tm[2 P 3/2] - \orb[3d ] \tm[2 D 3/2,5/2] &  & 1.399e+01 &  5.594e+01 &  1.893e+02 &  3.883e+01 \\
    \ion[Si iv]  &   1393.780 & \orb[3s ] \tm[2 S 1/2] - \orb[3p ] \tm[2 P 3/2] &  & 4.358e+02 &  3.648e+03 &  1.580e+04 &  1.468e+03 \\
    \ion[Si iv]  &   1402.770 & \orb[3s ] \tm[2 S 1/2] - \orb[3p ] \tm[2 P 1/2] &  & 2.270e+02 &  1.785e+03 &  7.347e+03 &  7.370e+02 \\
   \ion[Si vii]  &   1049.250 & \orb[2s 2]\orb[2p 4] \tm[3 P 1] - \orb[2s 2]\orb[2p 4] \tm[1 S 0] & 5.76 & 2.474e+00 &  2.128e+00 &  3.247e+00 &  1.895e+01 \\
    \ion[Si ix]  &    676.490 & \orb[2s 2]\orb[2p 2] \tm[3 P 1] - \orb[2s2p 3] \tm[5 S 2] & 6.06 & 6.827e-01 &  1.047e+00 &  3.022e+00 &  1.750e+00 \\
    \ion[S iii]  &   1015.500 & \orb[3s 2]\orb[3p 2] \tm[3 P 1] - \orb[3s3p 3] \orb[3 P 0,1] & 4.66 & 6.887e-01 &  2.669e+00 &  8.193e+00 &  2.039e+00 \\
    \ion[S iii]  &   1015.770 & \orb[3s 2]\orb[3p 2] \tm[3 P 1] - \orb[3s3p 3] \orb[3 P 2] &  & 5.016e-01 &  2.445e+00 &  6.699e+00 &  1.622e+00 \\
    \ion[S iii]  &   1021.080 & \orb[3s 2]\orb[3p 2] \tm[3 P 2] - \orb[3s3p 3] \orb[3 P 1] &  & 4.251e-01 &  2.649e+00 &  6.981e+00 &  1.149e+00 \\
    \ion[S iii]  &   1021.300 & \orb[3s 2]\orb[3p 2] \tm[3 P 2] - \orb[3s3p 3] \orb[3 P 2] &  & 1.192e+00 &  4.568e+00 &  2.288e+01 &  2.535e+00 \\
    \ion[S iii]  &   1200.990 & \orb[3s 2]\orb[3p 2] \tm[3 P 2] - \orb[3s3p 3] \orb[3 D 3] &  & 5.086e+00 &  1.990e+01 &  1.845e+02 &  1.516e+01 \\
     \ion[S iv]  &    744.910 & \orb[3s 2]\orb[3p ] \tm[2 P 1/2] - \orb[3s3p 2] \tm[2 P 3/2] & 4.99 & 5.552e-01 &  2.153e+00 &  1.169e+01 &  5.060e+00 \\
     \ion[S iv]  &    748.400 & \orb[3s 2]\orb[3p ] \tm[2 P 1/2] - \orb[3s3p 2] \tm[2 P 1/2] &  & 1.339e+00 &  3.940e+00 &  2.065e+01 &  1.176e+01 \\
     \ion[S iv]  &    750.220 & \orb[3s 2]\orb[3p ] \tm[2 P 3/2] - \orb[3s3p 2] \tm[2 P 3/2] &  & 2.766e+00 &  8.431e+00 &  5.318e+01 &  2.622e+01 \\
     \ion[S iv]  &    753.740 & \orb[3s 2]\orb[3p ] \tm[2 P 3/2] - \orb[3s3p 2] \tm[2 P 1/2] &  & 8.233e-01 &  2.544e+00 &  1.245e+01 &  6.416e+00 \\
     \ion[S iv]  &    815.950 & \orb[3s 2]\orb[3p ] \tm[2 P 3/2] - \orb[3s3p 2] \tm[2 S 1/2] &  & 1.936e+00 &  6.155e+00 &  2.438e+01 &  1.044e+01 \\
      \ion[S v]  &    786.470 & \orb[3s 2] \tm[1 S 0] - \orb[3s3p ] \tm[1 P 1] & 5.19 & 2.781e+01 &  4.254e+01 &  2.257e+02 &  5.116e+02 \\
     \ion[S vi]  &    706.500 & \orb[3p ] \tm[2 P 1/2] - \orb[3d ] \tm[2 D 3/2] & 5.29 & 1.020e+00 &  1.667e+00 &  2.746e+00 &  5.460e+00 \\
     \ion[Fe x]  &   1463.380 & \orb[3s 2]\orb[3p 4](\tm[3 P ])\orb[3d ] \tm[4 F 9/2] - \orb[3s 2]\orb[3p 4](\tm[1 D ])\orb[3d ] \tm[2 F 7/2] & 5.98 & 5.699e+01 &  2.279e+02 &  4.276e+02 &  1.147e+02 \\
   \ion[Fe xii]  &   1241.950 & \orb[3s 2]\orb[3p 3] \tm[4 S 3/2] - \orb[3s 2]\orb[3p 3] \tm[2 P 3/2] & 6.13 & 3.325e+00 &  2.596e+01 &  9.015e+01 &  1.098e+01 \\
   \ion[Fe xii]  &   1349.430 & \orb[3s 2]\orb[3p 3] \tm[4 S 3/2] - \orb[3s 2]\orb[3p 3] \tm[2 P 1/2] &  & 1.605e+00 &  1.551e+01 &  5.158e+01 &  8.580e+00 \\
 & \\
\hline
\enddata
\end{deluxetable}

 \begin{deluxetable}{l l c l@{~--~}l r r c }
\tablewidth{0pt} \tabletypesize{\scriptsize}
 \tablecaption{List of ``peculiar" lines identified in SUMER spectra of the sunspot.
$A_{ki}$ is the Einstein coefficient taken from NIST Atomic Spectral
Database (Ralchenko et al.,2008) and Chianti Database (Dere et al.,
1997;Landi et al., 2006). If available, unidentified lines are
characterized by the temperature classification of Feldman et
al.(1997) (a):$T_e<3\times10^5$~K, (b):$T_e\approx3\times10^5$~K,
(c):$T_e\approx4\times10^5$~K,
(d):$6\times10^5~\rm{K}<T_e<9\times10^5$~K,
(e):$T_e\approx1.4\times10^6$~K, (f):$T_e\approx1.8\times10^6$~K.
The line intensities are given in the unit of $\rm{mW~sr^{-1}
m^{-2}}$. References in last column: (0) Curdt et al.(2001), (1)
Ekberg \& Feldman (2003), (2) Sandlin et al.(1986) (3) New
identification. \label{table6}} \tablehead{\colhead{$\lambda_{obs}$
({\AA})} & \colhead{Line} & \colhead{Intensity} &
\multicolumn{2}{c}{Transition} & \colhead{$A_{ki}~(s^{-1})$}
&\colhead{Also~seen~in} & \colhead{Reference} } \tablecolumns{7}
\startdata
  668.41 &    (e)     & 0.31800 & \multicolumn{2}{c}{}   &     & streamer  & 0 \\
  669.01 & \ion[Al ix]& 0.13913 & \orb[2s 2]\orb[2p ]~\tm[2 P 1/2]& \orb[2s$\,$2p 2]~\tm[4 P 5/2]  & 5.19e-02   & streamer  & 0 \\
  670.02 & \ion[Al x]& 0.23453 & \orb[2s ]\orb[2p ]~\tm[1 P 1]& \orb[2p 2]~\tm[1 D 2]  & 1.57e-03   & streamer  & 0 \\
  672.55 &   ?  & 0.27117 &  \multicolumn{2}{c}{}   &  &   & 0 \\
  674.20 &   ?  & 0.34224 &  \multicolumn{2}{c}{}   &  &   & 0 \\
  679.78 & \ion[Mg viii]&  0.67506 & \orb[2s ]\orb[2p 2]~\tm[2 P 1/2]& \orb[2p 3]~\tm[2 D 3/2] & 6.81e+08 & weak in QS & 0 \\
  680.39 & \ion[Al ix]& 2.50337 & \orb[2s 2]\orb[2p ]~\tm[2 P 1/2]& \orb[2s ]\orb[2p 2]~\tm[4 P 3/2] & 3.84e+03 & weak in QS & 0 \\
  684.05 & \ion[Na vi]  &  0.37180 & \orb[2s ]\orb[2p 3]~\tm[3 D 3]& \orb[2s ]\orb[2p 3]~\tm[1 P 1] & 7.28e-03 & streamer  & 3 \\
  686.43 & \ion[Fe viii]& 2.02066 & \orb[3p 5]\orb[3d 2]~\tm[2 P 3/2]& \orb[3p 6]\orb[4d ]~\tm[2 D 5/2]& 9.71e-01 & & 1 \\
  688.72 & \ion[Fe viii]& 0.17972 & \orb[3p 5]\orb[3d 2 ]~\tm[2 P 3/2]& \orb[3p 6]\orb[4d ]~\tm[2 D 3/2]& 1.08e-01 & & 1 \\
  689.61 &\ion[Mg viii]& 0.81771 & \orb[2s ]\orb[2p 2]~\tm[2 P 3/2]& \orb[2p 3]~\tm[2 D 5/2] & 7.51e+08 & weak in QS & 0 \\
  691.40 &\ion[Ca ix]& 5.52513 & \orb[3s 2]~\tm[1 S 0]& \orb[3s$\,$3p ]~\tm[3 P 1] & 6.05e+05 & corona & 0 \\
  693.33 &    ? & 0.37120 &  \multicolumn{2}{c}{}  & & & 0 \\
  694.70 & \ion[Si ix]& 6.12387 & \orb[2s 2]\orb[2p 2]~\tm[3 P 2]&\orb[2s$\,$2p 3]~\tm[5 S 2] & 1.38e+05 & streamer & 0 \\
  697.14 & \ion[Fe viii]& 4.56267 & \orb[3p 6]\orb[4p ]~\tm[2 P 1/2]& \orb[3p 6]\orb[4d ]~\tm[2 D 3/2] & 2.00e+00 & weak in QS & 1 \\
  707.72 &     (c)    & 0.26568 & \multicolumn{2}{c}{}&    &       & 0 \\
  708.44 &     (c)    & 0.29286 & \multicolumn{2}{c}{}&    &       & 0 \\
  709.21 & \ion[Ar v] & 0.40630 & \orb[3s 2]\orb[3p 2]~\tm[3 P 1]& \orb[3s$\,$3p 3]~\tm[3 P 2] & & & 0 \\
  712.89 & \ion[S vi] & 4.34920 & \orb[2p 6]\orb[3p ]~\tm[2 P 3/2]&\orb[2p 6]\orb[3d ]~\tm[2 D 5/2,3/2] & 8.10e+08 & weak in QS & 0 \\
  713.81 & \ion[Ar viii]& 4.92706 & \orb[2p 6]\orb[3s ]~\tm[2 S 1/2]&\orb[2p 6]\orb[3p ]~\tm[2 P 1/2]& 2.40e+09 & weak in QS & 0 \\
  714.79 &     ?      & 0.22144 & \multicolumn{2}{c}{}&    &       & 0 \\
  715.65 & \ion[Ar v] & 0.81987 & \orb[3s 2]\orb[3p 2]~\tm[3 P 2]& \orb[3s$\,$3p 3]~\tm[3 P 1]&  & weak in QS & 0\\
  717.01 &     ? & 0.53082 & \multicolumn{2}{c}{}&    &  corona & 0 \\
  717.69 & \ion[Fe viii] ?& 0.31754 & \orb[3p 6]\orb[4p ]~\tm[2 P 3/2]& \orb[3p 6]\orb[4d ]~\tm[2 D 5/2]& & & 0 \\
  719.35 & \ion[S ix] & 0.24769 & \orb[2p 3]\orb[3p ](\tm[2 D ])\tm[3 F 4]&\orb[2p 3]\orb[3d ](\tm[3 D ])\tm[3 G 5]& 2.36e+09 & corona & 0 \\
  721.23 & \ion[Fe viii]& 8.42192 & \orb[3p 6]\orb[4p ]~\tm[2 P 1/2]& \orb[3p 6]\orb[4d ]~\tm[2 D 3/2]& 2.80e+00 & weak in QS & 1 \\
  723.75 & \ion[Fe viii]& 0.65898 & \orb[3p 6]\orb[4p ]~\tm[2 P 3/2]& \orb[3p 6]\orb[4d ]~\tm[2 D 3/2]& 3.10e-01 &  & 1 \\
  725.11 & \ion[Ar v] & 0.64130 & \orb[3s 2]\orb[3p 2]~\tm[1 D 2]& \orb[3s$\,$3p 3]~\tm[1 D 2]& & weak in QS  & 0 \\
  728.11 & \ion[Fe vii]& 0.41838 & \orb[3d ]\orb[4p ]~\tm[3 D 2]& \orb[3d ]\orb[4d ]~\tm[3 F 3]& & & 1 \\
  735.42 & \ion[Fe vii]& 0.39913 & \orb[3d ]\orb[4p ]~\tm[3 P 1]& \orb[3d ]\orb[4d ]~\tm[3 P 2]& & & 1 \\
  736.71 &   ?         & 0.78293 & \multicolumn{2}{c}{}&    &    &    0 \\
  738.87 & \ion[Fe vii]& 0.80524 & \orb[3d ]\orb[4p ]~\tm[3 P 2]& \orb[3d ]\orb[4d ]~\tm[3 P 2]& & & 1 \\
\t740.03 & \ion[Fe vii]& & \orb[3d ]\orb[4p ]~\tm[1 F 3]& \orb[3d ]\orb[4d ]~\tm[1 G 4]& & & 1 \\
\b740.11 & \ion[Ar viii]& & \orb[2p 6]\orb[4d ]~\tm[2 D 5/2]&\orb[2p 6]\orb[5p ]~\tm[2 P 3/2] & & & 0 \\
  740.79 & \ion[Fe vii]& 0.84513 & \orb[3d ]\orb[4p ]~\tm[3 D 3]& \orb[3d ]\orb[4d ]~\tm[3 F 4]& & & 1 \\
  741.14 & \ion[Fe vii]& 0.66444 & \orb[3d ]\orb[4p ]~\tm[3 P 1]& \orb[3d ]\orb[4d ]~\tm[3 P 0]& & & 1 \\
  745.38 & \ion[Fe vii]& 0.68352 & \orb[3d ]\orb[4p ]~\tm[3 F 2]& \orb[3d ]\orb[4d ]~\tm[3 F 2]& & & 1 \\
  747.37 & ?      & 0.55478 & \multicolumn{2}{c}{} &    &    &   0 \\
  749.01 & \ion[Fe vii]& 1.15036 & \orb[3d ]\orb[4p ]~\tm[3 F 3]& \orb[3d ]\orb[4d ]~\tm[3 F 3]& & & 1 \\
  750.58 &     (a)    & 1.38937 & \multicolumn{2}{c}{}&    & streamer   &    0 \\
  751.46 &     (a)    & 0.53180 & \multicolumn{2}{c}{}&    & streamer   &    0 \\
  752.51 & \ion[Fe viii]& 0.65775 & \orb[3p 5]\orb[3d 2]~\tm[2 P 1/2]& \orb[3p 6]\orb[4d ]~\tm[2 D 3/2]& & & 1 \\
  753.22 & \ion[Mg viii] & 1.08143 & \orb[2s 2]\orb[2p ]~\tm[2 P 1/2]& \orb[2s ]\orb[2p 2]~\tm[4 D 5/2] & 3.46e-02 & corona & 3 \\
  754.93 & \ion[Ar vi]& 4.15884 & \orb[3s 2]\orb[3p ]~\tm[2 P 1/2]& \orb[3s$\,$3p 2]~\tm[2 D 3/2]  & 2.80e+09 & & 0 \\
  756.71 & \ion[Al viii]& 0.87528 & \orb[2s 2]\orb[2p 2]~\tm[3 P 1]& \orb[2s$\,$2p 3]~\tm[5 S 2] & 3.00e+04 & & 0 \\
  757.15 & \ion[Fe vii]& 2.65446 & \orb[3d ]\orb[4p ]~\tm[3 F 4]& \orb[3d ]\orb[4d ]~\tm[3 F 4]& & &  1 \\
  762.65 & \ion[Mg viii]& 1.77679 & \orb[2s 2]\orb[2p ]~\tm[2 P 1/2]& \orb[2s$\,$2p 2]~\tm[4 P 3/2] & 1.71e+03 & corona & 0 \\
  767.07 & \ion[Ar vi]& 9.64480 & \orb[3s 2]\orb[3p ]~\tm[2 P 3/2]& \orb[3s$\,$3p 2]~\tm[2 D 5/2] & 3.30e+09  & weak in QS & 0 \\
  769.38 & \ion[Mg viii]& 6.67460 & \orb[2s 2]\orb[2p ]~\tm[2 P 1/2]&\orb[2s$\,$2p 2]~\tm[4 P 1/2] & 7.29e+04 & corona & 0 \\
  772.26 & \ion[Mg viii]& 43.7776 & \orb[2s 2]\orb[2p ]~\tm[2 P 3/2]&\orb[2s ]\orb[2p 2]~\tm[4 P 5/2]& 5.67e+04 & corona & 0 \\
  776.62 &   (b)        & 3.69754 & \multicolumn{2}{c}{}&    &  streamer  &   0 \\
  782.34 & \ion[Mg viii]& 7.38377 & \orb[2s 2]\orb[2p ]~\tm[2 P 3/2]& \orb[2s$\,$2p 2]~\tm[4 P 3/2] & 1.61e+04 & corona &   0 \\
  784.52 & \ion[Fe vii]& 2.02086 & \orb[3d$\,$4p ]~\tm[3 F 1]& \orb[3d$\,$4d ]~\tm[3 S 1]& & & 1 \\
  789.43 & \ion[Mg viii]& 5.96976 & \orb[2s 2]\orb[2p ]~\tm[2 P 3/2]& \orb[2s$\,$2p 2]~\tm[4 P 1/2]& 5.96e+04 & corona & 0 \\
  789.78 & \ion[Na viii]& 4.45572 & \orb[2s 2]~\tm[1 S 0]& \orb[2s$\,$2p ]~\tm[3 P 1] & 3.63e+04 & & 0 \\
  792.77 & \ion[Fe vii]& 3.80994 & \orb[3d ]\orb[4p ]~\tm[3 F 2]& \orb[3d ]\orb[4d ]~\tm[3 G 3]& & & 1 \\
  793.75 &     ?      & 1.54882 & \multicolumn{2}{c}{}&    &    &   3 \\
  794.25 &     ?      & 1.28655 & \multicolumn{2}{c}{}&    &    &   3 \\
  795.23 &     (a)    & 1.54914 & \multicolumn{2}{c}{}&    & streamer &   0 \\
  799.11 & \ion[Fe vii]& 1.73620 & \orb[3d ]\orb[4p ]~\tm[3 D 3]& \orb[3d ]\orb[4d ]~\tm[3 D 3]& & & 1 \\
  800.68 & \ion[Cl vii]& 2.99016 & \orb[2p 6]\orb[3s ]~\tm[2 S 1/2]&\orb[2p 6]\orb[3p ]~\tm[2 P 3/2]& 2.17e+09 & & 0 \\
  801.69 & \ion[Fe vii]& 5.08175 & \orb[3d ]\orb[4p ]~\tm[3 F 3]& \orb[3d ]\orb[4d ]~\tm[3 G 3]& & & 1 \\
  803.46 &     (e)    & 3.26169 & \multicolumn{2}{c}{}&    &   streamer & 0 \\
  804.16 & \ion[Fe vii]& 8.47300 & \orb[3d ]\orb[4p ]~\tm[3 F 4]& \orb[3d ]\orb[4d ]~\tm[3 G 5]& & & 1 \\
  811.55 & \ion[Mn vii]& 1.67024 & \orb[3p 6]\orb[4p ]~\tm[2 P 3/2]& \orb[3p 6]\orb[4d ]~\tm[2 D 5/2]& & & 3 \\
  813.01 & \ion[Cl vii]& 2.43765 & \orb[2p 6]\orb[3s ]~\tm[2 S 1/2]&\orb[2p 6]\orb[3p ]~\tm[2 P 1/2]& 2.11e+09 & & 0 \\
  821.23 & \ion[Ca ix] & 3.87217 & \orb[3s$\,$3p ]~\tm[1 P 1]& \orb[3p 2]~\tm[1 D 2]& 6.79e+08 &   & 0 \\
  827.06 & \ion[Ar v]  & 2.10911 & \orb[3s 2]\orb[3p 2]~\tm[3 P 1]& \orb[3s$\,$3p 3]~\tm[3 D 2]& 2.20e+09 &  & 0 \\
  849.29 & \ion[S v]   & 4.01087 & \orb[3s$\,$3p ]~\tm[3 P 1]& \orb[3p 2]~\tm[3 P 2] & 1.07e+09 &  & 0 \\
  852.17 & \ion[S v]   & 5.69634 & \orb[3s$\,$3p ]~\tm[3 P 0]& \orb[3p 2]~\tm[3 P 1] & 1.50e+09 &   & 0 \\
\t854.71 & \ion[Mg vii]&  & \orb[2s 2]\orb[2p 2]~\tm[3 P 1]& \orb[2s$\,$2p 3]~\tm[5 S 2]& 1.14e+04 & & 0 \\
\b854.80 & \ion[S v]   &  & \orb[3s$\,$3p ]~\tm[3 P 2]& \orb[3p 2]~\tm[3 P 2] & 3.13e+09 & & 0 \\
  857.82 & \ion[S v]   & 2.94460 & \orb[3s$\,$3p ]~\tm[3 P 1]& \orb[3p 2]~\tm[3 P 0] & 4.14e+09 & & 0 \\
  860.48 & \ion[S v]   & 6.74638 & \orb[3s$\,$3p ]~\tm[3 P 2]& \orb[3p 2]~\tm[3 P 1] & 1.71e+09 & & 0 \\
  868.13 & \ion[Mg vii]& 44.0762 & \orb[2s 2]\orb[2p 2]~\tm[3 P 2]& \orb[2s$\,$2p 3]~\tm[5 S 2] & 2.90e-02 & & 0 \\
  872.12 & \ion[Na vii]& 9.19926 & \orb[2s 2]\orb[2p ]~\tm[2 P 3/2]& \orb[2s$\,$2p 2]~\tm[4 P 5/2]& 2.84e+04 & & 0 \\
  873.78 & \ion[Si v] ?  & 6.17296 & \orb[2p 5]\orb[3s ]~\tm[1 P 1]& \orb[2p 5]\orb[3p ]~\tm[1 S 0]& 5.06e+09 & streamer & 3 \\
  877.92 & \ion[Ar vii] ? & 7.16439 & \orb[3s 2]~\tm[1 S 0]& \orb[3s$\,$3p ]~\tm[3 P 1]    &   &  & 0 \\
  880.33 & \ion[Na vii]& 3.85727 & \orb[2s 2]\orb[2p ]~\tm[2 P 3/2]& \orb[2s$\,$2p 2]~\tm[4 P 3/2]& 7.77e+03 & & 0 \\
  885.33 &     (b)     & 5.53839 & \multicolumn{2}{c}{}&      & streamer & 0 \\
  895.17 & \ion[Ne vii]& 80.3477 & \orb[2s 2]~\tm[1 S 0]& \orb[2s$\,$2p ]~\tm[3 P 1] & 1.32e+04 & weak in QS & 0 \\
  905.02 & \ion[Si vii]& 8.35475 & \orb[2p 3]\orb[3p ]~\tm[5 P 2]& \orb[2p 3]\orb[3d ]~\tm[5 D 2,3] & 1.22e+09 & & 0 \\
  944.34 & \ion[Si viii]& 33.6785 & \orb[2s 2]\orb[2p 3]~\tm[4 S 3/2]&\orb[2s 2]\orb[2p 3]~\tm[2 P 3/2]& 7.23e+01 & weak in QS & 0 \\
  957.11 & \ion[O vi]? & 1.59753 & \multicolumn{2}{c}{} &   & streamer &  0 \\
  965.28 &     (d)  & 1.18173 & \multicolumn{2}{c}{}  &    & streamer & 0 \\
  966.68 &     ?    & 0.92318 & \multicolumn{2}{c}{} &     &   & 0 \\
  975.85 &     (d)  & 0.96417 & \multicolumn{2}{c}{} &     & streamer &  0 \\
  997.21 & \ion[Ne vi] & 11.5087 & \orb[2s 2]\orb[2p ]~\tm[2 P 1/2]& \orb[2s$\,$2p 2]~\tm[4 P 1/2]& 1.18e+04 & & 0 \\
  998.05 &     ?    & 2.18132 & \multicolumn{2}{c}{} &     &   & 0 \\
  999.25 & \ion[Ne vi] & 62.3252 & \orb[2s 2]\orb[2p ]~\tm[2 P 3/2]& \orb[2s$\,$2p 2]~\tm[4 P 5/2]& 9.08e+03 & weak in QS &0 \\
 1005.79 & \ion[Ne vi] & 32.5323 & \orb[2s 2]\orb[2p ]~\tm[2 P 3/2]& \orb[2s$\,$2p 2]~\tm[4 P 3/2]& 2.92e+03  &   & 0 \\
 1010.25 & \ion[Ne vi] & 12.4947 & \orb[2s 2]\orb[2p ]~\tm[2 P 3/2]& \orb[2s$\,$2p 2]~\tm[4 P 1/2]& 1.06e+04  &   & 0 \\
 1011.61 &     ?       & 1.31622 & \multicolumn{2}{c}{}&     &    & 0 \\
 1012.65 &     ?       & 1.96463 & \multicolumn{2}{c}{}&     &    & 0 \\
 1013.88 &     ?       & 1.26811 & \multicolumn{2}{c}{}&     &    & 0 \\
 1028.95 &     (e)     & 3.54614 & \multicolumn{2}{c}{}&     & streamer & 0 \\
 1049.25 & \ion[Si vii]& 12.4464 & \orb[2s 2]\orb[2p 4]~\tm[3 P 1]& \orb[2s 2]\orb[2p 4]~\tm[1 S 0]& 1.39e+02 &  & 0 \\
 1051.60 & \ion[O iii]/2& 33.8099 & \orb[2s 2]\orb[2p 2]~\tm[1 D 2]& \orb[2s$\,$2p 3]~\tm[1 P 1] & 9.60e+09 & & 0 \\
 1053.87 & \ion[Al vii]& 4.47774 & \orb[2s 2]\orb[2p 3]~\tm[4 S 3/2]& \orb[2s 2]\orb[2p 3]~\tm[2 P 3/2]& 3.13e+01 & & 0 \\
 1056.81& \ion[Al vii]& 2.29385 & \orb[2s 2]\orb[2p 3]~\tm[4 S 3/2]& \orb[2s 2]\orb[2p 3]~\tm[2 P 1/2]& 1.28e+01 & weak in QS & 0 \\
 1057.81 & \ion[Al viii]& 1.73163 & \orb[2s 2]\orb[2p 2]~\tm[3 P 1]& \orb[2s 2]\orb[2p 2]~\tm[1 S 0]& 9.05e+01 & & 0 \\
 1067.83 & \ion[O iv]  & 3.86110 & \orb[2s 2]\orb[3d ]~\tm[2 D 5/2]& \orb[2s 2]\orb[4f ]~\tm[2 F 7/2]& 3.33e+09 & & 0 \\
\t1087.86& \ion[Ne iv]/2& & \orb[2s 2]\orb[2p 3]~\tm[3 S 3/2]& \orb[2s ]\orb[2p 4]~\tm[4 P 5/2]& 2.50e+09 &  & 3\\
\b1087.86& \ion[Fe vii]& & \orb[3d ]\orb[4s ]~\tm[3 D 2]& \orb[3d ]\orb[4p ]~\tm[3 P 1] & & & 1 \\
 1095.37 & \ion[Fe vii]& 3.09757 & \orb[3d ]\orb[4s ]~\tm[3 D 3]& \orb[3d ]\orb[4p ]~\tm[3 P 2]& & & 1 \\
 1098.92 &   ?     & 2.72720 & \multicolumn{2}{c}{}  &  &  & 0\\
 1106.70 & \ion[O iv]/2& 169.437 & \orb[2s 2]\orb[2p ]~\tm[2 P 1/2]&\orb[2s ]\orb[2p 2]~\tm[2 P 3/2]& 1.22e+09 & weak in QS & 0 \\
 1108.16 & \ion[O iv]/2& 195.290 & \orb[2s 2]\orb[2p ]~\tm[2 P 1/2]& \orb[2s$\,$2p 2]~\tm[2 P 1/2] & 4.86e+09 & weak in QS & 0 \\
 1109.12 & \ion[O iv]/2& 685.028 & \orb[2s 2]\orb[2p ]~\tm[2 P 3/2]& \orb[2s$\,$2p 2]~\tm[2 P 3/2]& 6.06e+09 & weak in QS & 0 \\
 1110.52 & \ion[O iv]/2& 132.044 & \orb[2s 2]\orb[2p ]~\tm[2 P 3/2]& \orb[2s$\,$2p 2]~\tm[2 P 1/2]& 2.41e+09 & weak in QS & 0 \\
 1115.52 & \ion[Ca x]/2& 49.6308 & \orb[2p 6]\orb[3s ]~\tm[2 S 1/2]& \orb[2p 6]\orb[3p ]~\tm[2 P 3/2]& 3.50e+09 &  & 0 \\
 1119.10 &  H$_2$      & 113.679 & \multicolumn{2}{l}{{\it 1 - 3}$\;Q3$ \hspace{3mm}{\it (C-X)}} &    &   & 0 \\
 1125.66 & \ion[Ne vi]/2 & 343.100 & \orb[2s 2]\orb[2p ]~\tm[2 P 3/2]& \orb[2s$\,$2p 2]~\tm[2 D 5/2,3/2] & 1.17e+09 & weak in QS & 0 \\
 1128.35 & \ion[Si iv]& 36.3423 & \orb[2p 6]\orb[3p ]~\tm[2 P 3/2]& \orb[2p 6]\orb[3d ]~\tm[2 D 5/2,3/2]& 2.53e+09 & weak in QS & 0 \\
 1136.56 & \ion[Ne v] & 19.1372 & \orb[2s 2]\orb[2p 2]~\tm[3 P 1]& \orb[2s$\,$2p 3]~\tm[5 S 2] & 2.23e+03 & weak in QS & 0 \\
 1136.82 & \ion[Ne v]/2 & 47.5348 & \orb[2s 2]\orb[2p 2]~\tm[3 P 0]& \orb[2s$\,$2p 3]~\tm[3 D 1]& 7.11e+08 & & 0 \\
 1141.45 & \ion[Fe vii]& 5.09396 & \orb[3d ]\orb[4s ]~\tm[3 D 3]& \orb[3d ]\orb[4p ]~\tm[3 F 4]& & & 1 \\
 1144.28 & \ion[Ne v]/2 & 15.2854 & \orb[2s 2]\orb[2p 2]~\tm[3 P 2]& \orb[2s$\,$2p 3]~\tm[3 D 1]& 2.88e+07 & & 0 \\
 1144.74 & \ion[Ne v]/2 & 101.738 & \orb[2s 2]\orb[2p 2]~\tm[3 P 2]& \orb[2s$\,$2p 3]~\tm[3 D 3]& 2.76e+08 & & 0 \\
 1145.66 & \ion[Ne v] & 33.2632 & \orb[2s 2]\orb[2p 2]~\tm[3 P 2]& \orb[2s$\,$2p 3]~\tm[5 S 2]& 6.38e+03 & weak in QS & 0 \\
 1148.06 & \ion[Ca x]/2& 23.1761 & \orb[2p 6]\orb[3s ]~\tm[2 S 1/2]&\orb[2p 6]\orb[3p ]~\tm[2 P 1/2]& 3.20e+09 &  & 0 \\
 1149.59 & (d) & 1.34829 & \multicolumn{2}{c}{} & & & 3\\
 1155.01 & \ion[Fe vii]& 0.52653 & \orb[3d ]\orb[4s ]~\tm[3 D 2]& \orb[3d ]\orb[4p ]~\tm[3 F 3] & & & 1 \\
 1163.85 & H$_2$ & 180.646 & \multicolumn{2}{l}{{\it 1 - 4}$\;Q3$ \hspace{3mm}{\it (C-X)}} & & & 0 \\
 1165.74 & \ion[Ca viii]/2 & 41.3746 & \orb[3s 2]\orb[3p ]~\tm[2 P 1/2]& \orb[3s$\,$3p 2]~\tm[2 D 3/2]& 4.10e+09 & & 0 \\
 1166.22 & \ion[Fe vii]& 2.91614 & \orb[3d ]\orb[4s ]~\tm[3 D 1]& \orb[3d ]\orb[4p ]~\tm[3 F 2]&  &  & 1 \\
 1170.18 & ? & 10.1514 & \multicolumn{2}{c}{} & & & 0 \\
 1171.60 & \ion[Ar vii]/2 & 84.1832 & \orb[3s 2]~\tm[1 S 0]& \orb[3s$\,$3p ]~\tm[1 P 1]& 7.83e+09 &    & 0 \\
 1189.84 & \ion[Mg vii]& 27.2145 & \orb[2s 2]\orb[2p 2]~\tm[3 P 1]& \orb[2s 2]\orb[2p 2]~\tm[1 S 0]& 3.62e+01 & & 0 \\
 1190.12 & \ion[Mg vi]& 123.208 & \orb[2s 2]\orb[2p 3]~\tm[4 S 3/2]& \orb[2s 2]\orb[2p 3]~\tm[2 P 3/2]& 1.21e+01 & & 0 \\
 1191.68 & \ion[Mg vi]& 30.3776 & \orb[2s 2]\orb[2p 3]~\tm[4 S 3/2]& \orb[2s 2]\orb[2p 3]~\tm[2 P 1/2]& 4.91e+00 & & 0 \\
 1199.21 & \ion[S v]  & 91.5734 & \orb[3s 2]~\tm[1 S 0]& \orb[3s$\,$3p ]~\tm[3 P 1]  & 1.23e+05 & weak in QS & 0 \\
 1209.04 &  H$_2$     & 121.340 & \multicolumn{2}{l}{{\it 1 - 5}$\;Q3$ \hspace{3mm}{\it (C-X)}}&  &  & 0 \\
 1218.34 & \ion[O v]  & 2085.46 & \orb[2s 2]~\tm[1 S 0]& \orb[2s$\,$2p ]~\tm[3 P 1]& 2.34e+03 &  & 0 \\
 1251.48 & \ion[Si v] & 7.48967 & \orb[2s 2]\orb[2p 5]\orb[3p ]~\tm[3 P 2]& \orb[2s 2]\orb[2p 5]\orb[3p ]~\tm[3 D 3]& 9.59e+08 & & 3 \\
 1251.76 & \ion[Si v] & 4.77757 & \orb[2s 2]\orb[2p 5]\orb[3p ]~\tm[1 P 1]& \orb[2s 2]\orb[2p 5]\orb[3p ]~\tm[3 P 1]& 3.96e+08 & & 3 \\
 1254.11 &  H$_2$     & 17.6018 & \multicolumn{2}{l}{{\it 1 - 6}$\;Q3$ \hspace{3mm}{\it (C-X)}}& & & 0 \\
 1257.24 &  H$_2$     & 2.94243 & \multicolumn{2}{l}{{\it 1 - 3}$\;R3$ \hspace{3mm}{\it (X-B)}}& & & 2 \\
 1267.76 & \ion[Ca vi]/2& 14.2663 & \orb[3s 2]\orb[3p 3]~\tm[4 S 3/2] &\orb[3s$\,$3p 4]~\tm[4 P 3/2]& 5.00e+09 & & 0 \\
 1283.92 & \ion[Ca vi]/2& 29.4227 & \orb[3s 2]\orb[3p 3]~\tm[4 S 3/2] &\orb[3s$\,$3p 4]~\tm[4 P 5/2]& 4.70e+09 & & 0 \\
 1285.47 & \ion[Si v] & 2.21361& \orb[2s 2]\orb[2p 5]\orb[3s ]~\tm[3 P 0]& \orb[2s 2]\orb[2p 5]\orb[3p ]~\tm[3 D 1] & 2.67e+08 &      & 3 \\
 1285.68 &  H$_2$   & 1.85936 & \multicolumn{2}{l}{{\it 1 - 3}$\;P7$ \hspace{3mm}{\it (X-B)}}& & & 0 \\
 1286.20& \ion[Ca v]/2& 4.63060 & \orb[3s 2]\orb[3p 4]~\tm[3 P 1]& \orb[3s$\,$3p 5]~\tm[3 P 0]& 9.10e+08 & weak in QS & 0 \\
 1286.43 &  H$_2$   & 1.19117 &  \multicolumn{2}{l}{{\it 3 - 4}$\;P5$ \hspace{3mm}{\it (X-B)}} & & & 0 \\
 1293.14 & \ion[Ca v]/2& 4.54793 & \orb[3s 2]\orb[3p 4]~\tm[3 P 2]& \orb[3s$\,$3p 5]~\tm[3 P 2]& 6.90e+08 & & 0 \\
 1293.88 & \ion[Mg v] & 0.69026 & \orb[2s 2]\orb[2p 4]~\tm[3 P 2]& \orb[2s 2]\orb[2p 4]~\tm[1 S 0] & & & 3 \\
 1314.67 & \ion[S iv]/2& 20.7701 & \orb[3s 2]\orb[3p ]~\tm[2 P 1/2]&\orb[3s 2]\orb[3d ]~\tm[2 D 3/2] & 8.69e+09 & weak in QS & 0\\
 1317.84 &  ?          & 1.43195 & \multicolumn{2}{c}{}  &   &   & 0 \\
 1319.76 & \ion[S v]/2& 9.49761 & \orb[3s$\,$3p ]~\tm[3 P 1]& \orb[3s$\,$3d ]~\tm[3 D 2,1] & 4.87e+09 & weak in QS & 0 \\
 1322.84 & \ion[S iv]/2& 34.1715 & \orb[3s 2]\orb[3p ]~\tm[2 P 3/2]& \orb[3s 2]\orb[3d ]~\tm[2 D 5/2,3/2]& 1.03e+10 & weak in QS & 0 \\
 1324.59 & \ion[Mg v] & 30.8036 & \orb[2s 2]\orb[2p 4]~\tm[3 P 1]&\orb[2s 2]\orb[2p 4]~\tm[1 S 0]& 2.15e+01 & & 0 \\
 1326.36 & \ion[S v]/2& 16.7918 & \orb[3s$\,$3p ]~\tm[3 P 2]&\orb[3s$\,$3d ]~\tm[3 D 3,2]  & 6.39e+09 & & 0 \\
 1338.57 &  H$_2$     & 3.96306 &  \multicolumn{2}{l}{{\it 0 - 4}$\;P2$ \hspace{3mm}{\it (X-B)}}  & & & 0 \\
 1339.42 & \ion[Al v]/2 & 3.40516 & \orb[2s 2]\orb[2p 4](\tm[3 P ])\orb[3d ]~\tm[4 F 7/2]&\orb[2s 2]\orb[2p 4](\tm[3 P 2])\orb[4f ]~\tm[2 [4] 7/2] & & & 3 \\
 1342.88 & \ion[Cl vi]/2& 6.51297 & \orb[2p 6]\orb[3s 2]~\tm[1 S 0] &\orb[3s$\,$3p ]~\tm[1 P 1] & 6.32e+09  &    & 3\\
 1343.64 &   ?     & 4.24962 & \multicolumn{2}{c}{} &     &     &    0 \\
 1347.06 &   ?     & 4.15148 & \multicolumn{2}{c}{} &     &     &    0 \\
 1348.79 &   ?     & 2.67438 & \multicolumn{2}{c}{} &     &     &    0 \\
 1349.43 & \ion[Fe xii]& 10.4184 & \orb[3s 2]\orb[3p 3]~\tm[4 S 3/2]&\orb[3s 2]\orb[3p 3]~\tm[2 P 1/2] & 1.93e+02 &     &    0 \\
 1356.47 &  H$_2$  & 2.37708 &  \multicolumn{2}{l}{{\it 0 - 4}$\;R7$ \hspace{3mm}{\it (X-B)}} & & & 0 \\
 1360.78 & \ion[Al ix]/2 & 8.19409 & \orb[2s 2]\orb[2p ]~\tm[2 P 1/2]&\orb[2s$\,$2p 2]~\tm[4 P 3/2] & 3.57e+03 &  weak in QS &    0 \\
 1364.87 & \ion[Na v]  & 1.65732 & \orb[2s 2]\orb[2p 3]~\tm[4 S 3/2]&\orb[2s 2]\orb[2p 3]~\tm[2 P 3/2]& 4.16e+00 & & 0\\
 1365.51 & \ion[Na v]  & 2.25491 & \orb[2s 2]\orb[2p 3]~\tm[4 S 3/2]&\orb[2s 2]\orb[2p 3]~\tm[2 P 1/2]& 4.16e+00 & & 0\\
 1367.12 & \ion[S iii] ?& 1.58489 & \orb[3s 2]\orb[3p 2]~\tm[1 D 2]&\orb[3s 2]\orb[3p ]\orb[3d ]~\tm[1 F 3]&    &    & 3 \\
 1371.32 & \ion[O v]   & 164.349 & \orb[2s$\,$2p ]~\tm[1 P 1]& \orb[2p 2]~\tm[1 D 2] & 3.36e+08 &  & 0 \\
 1372.89 & \ion[Fe viii]/2& 3.83338 & \orb[3p 5]\orb[3d 2]~\tm[2 P 3/2]& \orb[3p 6]\orb[4d ]~\tm[2 D 5/2] & 2.73e+09 &  & 1 \\
 1382.88 & \ion[Ca ix]/2& 12.1360 & \orb[3s 2]~\tm[1 S 0]& \orb[3s$\,$3p ]~\tm[3 P 1] & 6.05e+05 & corona & 0 \\
 1385.74 &   ?     & 1.51274 & \multicolumn{2}{c}{} &     &     &    0 \\
 1389.44 & \ion[Si ix]/2& 5.29982 & \orb[2s 2]\orb[2p 2]~\tm[3 P 2]& \orb[2s$\,$2p 3]~\tm[5 S 2]& 1.38e+05 & corona & 0 \\
 1394.28 & \ion[Fe viii]/2& 38.5647 & \orb[3p 6]\orb[4p ]~\tm[2 P 1/2]& \orb[3p 6]\orb[4d ]~\tm[2 D 3/2] & 8.21e+09 &  & 1 \\
 1397.22 & \ion[O iv]  & 20.9203 & \orb[2s 2]\orb[2p ]~\tm[2 P 1/2]& \orb[2s$\,$2p 2]~\tm[4 P 3/2]& 3.75e+01 &  & 0 \\
 1398.06 & \ion[S iv]  & 2.50008 & \orb[3s 2]\orb[3p ]~\tm[2 P 1/2]& \orb[3s$\,$3p 2]~\tm[4 P 3/2]& 1.05e+03 &  & 0 \\
 1399.77 & \ion[O iv]  & 91.5729 & \orb[2s 2]\orb[2p ]~\tm[2 P 1/2]& \orb[2s$\,$2p 2]~\tm[4 P 1/2]& 1.47e+03 & weak in QS & 0 \\
 1400.52 & \ion[Ar viii]/2& 23.3320 & \orb[2p 6]\orb[3s ]~\tm[2 S 1/2]& \orb[2p 6]\orb[3p ]~\tm[2 P 3/2]& 2.73e+09 & corona & 0 \\
 1401.16 & \ion[O iv]  & 437.363 & \orb[2s 2]\orb[2p ]~\tm[2 P 3/2]& \orb[2s$\,$2p 2]~\tm[4 P 5/2]& 1.16e+03 & weak in QS & 0 \\
\t1404.79 & \ion[S iv]  & & \orb[3s 2]\orb[3p ]~\tm[2 P 1/2]& \orb[3s$\,$3p 2]~\tm[4 P 1/2]& 6.39e+04 & weak in QS & 0 \\
\b1404.82 & \ion[O iv]  & & \orb[2s 2]\orb[2p ]~\tm[2 P 3/2]& \orb[2s$\,$2p 2]~\tm[4 P 3/2]& 2.90e+02 & weak in QS & 0 \\
 1406.04 & \ion[S iv]  & 30.9620 & \orb[3s 2]\orb[3p ]~\tm[2 P 3/2]& \orb[3s$\,$3p 2]~\tm[4 P 5/2]& 5.13e+04 & & 0 \\
 1407.39 & \ion[O iv]  & 80.3204 & \orb[2s 2]\orb[2p ]~\tm[2 P 3/2]& \orb[2s$\,$2p 2]~\tm[4 P 1/2]& 1.45e+03 & & 0 \\
 1412.10 & \ion[Mg ix]/2& 26.5676 & \orb[2s 2]~\tm[1 S 0]& \orb[2s$\,$2p ]~\tm[3 P 1] & 9.04e+04 & & 0 \\
 1413.65 & ?           & 2.43612 & \multicolumn{2}{c}{} &     &     &    0 \\
 1416.93 & \ion[S iv]  & 19.6082 & \orb[3s 2]\orb[3p ]~\tm[2 P 3/2]& \orb[3s$\,$3p 2]~\tm[4 P 3/2]& 2.16e+04 &  & 0 \\
 1423.86 & \ion[S iv]  & 5.75110 & \orb[3s 2]\orb[3p ]~\tm[2 P 3/2]& \orb[3s$\,$3p 2]~\tm[4 P 1/2]& 4.72e+04 &  & 0 \\
 1425.42 & \ion[S vi]/2& 11.0943 & \orb[2p 6]\orb[3p ]~\tm[2 P 3/2]& \orb[2p 6]\orb[3d ]~\tm[2 D 5/2,3/2]& 5.20e+09 & & 0 \\
 1426.49 &  H$_2$      & 2.76212 & \multicolumn{2}{l}{{\it 0 - 5}$\;P7$ \hspace{3mm}{\it (X-B)}}   &  &  & 0 \\
 1427.68 & \ion[Ar viii]/2& 9.64131 & \orb[2p 6]\orb[3s ]~\tm[2 S 1/2]& \orb[2p 6]\orb[3p ]~\tm[2 P 1/2]& 2.40e+09 & & 0 \\
 1431.03 &  H$_2$      & 3.33829 & \multicolumn{2}{l}{{\it 1 - 6}$\;R3$ \hspace{3mm}{\it (X-B)}}   &     &     & 2 \\
 1431.32 & \ion[Ar v]/2& 2.33605 & \orb[3s 2]\orb[3p 2]~\tm[3 P 2]& \orb[3s$\,$3p 3]~\tm[3 P 2,1] & & & 0 \\
 1436.52 &   (e)/2     & 11.2925 & \multicolumn{2}{c}{}   &     &     & 0 \\
 1439.11 &  H$_2$?     & 2.37439 & \multicolumn{2}{c}{}   &     &     & 3 \\
 1442.54 & \ion[Fe viii]/2& 13.2009 & \orb[3p 6]\orb[4p ]~\tm[2 P 3/2]& \orb[3p 6]\orb[4d ]~\tm[2 D 5/2] & 2.80e+00 &  & 1 \\
 1444.09 &  CO         & 5.07571 & \multicolumn{2}{l}{{\it 5 - 1}$\;Q31$ \hspace{3mm}{\it (A-X)}}&     &     & 2 \\
 1445.77 & \ion[Si viii]& 16.3185 & \orb[2s 2]\orb[2p 3]~\tm[4 S 3/2]& \orb[2s 2]\orb[2p 3]~\tm[2 D 3/2]& 1.55e+00 &  & 0 \\
 1446.12 &  H$_2$      & 5.06184 & \multicolumn{2}{l}{{\it 1 - 6}$\;P5$ \hspace{3mm}{\it (X-B)}}& & & 3 \\
 1453.08 &  H$_2$      & 3.71507 & \multicolumn{2}{l}{{\it 0 - 5}$\;P10$ \hspace{3mm}{\it (X-B)}} &      &    & 2 \\
 1458.14 &  ?          & 3.33749 & \multicolumn{2}{c}{}   &      & weak in QS & 0 \\
 1477.77 & \ion[Fe vii]/2& 41.6691 & \orb[3d ]\orb[4p ]~\tm[3 P 2]& \orb[3d ]\orb[4d ]~\tm[3 P 2] & & & 1 \\
 1480.22& \ion[Ar viii]/2& 179.944 & \orb[2p 6]\orb[4d ]~\tm[2 D 5/2]&\orb[2p 6]\orb[5p ]~\tm[2 P 3/2]& & & 0 \\
 1482.36& \ion[Fe vii]/2& 37.2067 & \orb[3d ]\orb[4p ]~\tm[3 P 1]& \orb[3d ]\orb[4d ]~\tm[3 P 0] &  &  & 1 \\
 & \\
\hline
\enddata
\end{deluxetable}